\documentclass[useAMS,usenatbib]{mn2e}
\usepackage{epsfig}     % for figure inclusion
\usepackage{times,amsmath,amssymb}

\newcommand\mat[1]{\ensuremath{\bmath{#1}}}
\newcommand\vect[1]{\ensuremath{\bmath{#1}}}
\newcommand\av{\ensuremath{A_V}}
\newcommand\mv{\ensuremath{M_V}}

\newcommand\feh{\ensuremath{\mathrm{[M/H]}}}

\newcommand\afeh{\ensuremath{[\alpha/\mathrm{Fe}]}}

\newcommand\logg{\ensuremath{\log g}}

\newcommand\teff{\ensuremath{T_\mathrm{eff}}}

\newcommand\skgoal{\ensuremath{\sigma_{k,\mathrm{goal}}}}
\newcommand\skpost{\ensuremath{\sigma_{k,\mathrm{post}}}}
\newcommand\sikgoal{\ensuremath{\sigma_{ik,\mathrm{goal}}}}
\newcommand\sikpost{\ensuremath{\sigma_{ik,\mathrm{post}}}}
\newcommand\sikprior{\ensuremath{\sigma_{ik,\mathrm{prior}}}}
\newcommand\qnorm{\ensuremath{\widehat{Q}}}

\newcommand\vecphi{\ensuremath{\boldsymbol{\phi}}}

\newcommand{\Gaia}{\textit{Gaia}}

\title[A photometric system for {\Gaia}]
{The design and performance of the {\Gaia} photometric system}
\author[Jordi et al]{
C. Jordi$^{1,2}$\thanks{E-mail: carme@am.ub.es},
E. H{\o}g$^{3}$,
A.G.A. Brown$^{4}$,
L. Lindegren$^{5}$,
C.A.L. Bailer-Jones$^{6}$,
J.M. Carrasco$^{1}$, \newauthor
J. Knude$^{3}$,
V. Strai\v{z}ys$^{8}$,
J.H.J. de Bruijne$^{10}$,
J.-F. Claeskens$^{11}$,
R. Drimmel$^{12}$,  
F. Figueras$^{1,2}$, \newauthor
M. Grenon$^{7}$,
I. Kolka$^{13}$,
M.A.C. Perryman$^{10}$,
G. Tautvai{\v s}ien{\. e}$^{8}$,
V. Vansevi\v{c}ius$^{9}$,\newauthor
P.G. Willemsen$^{14}$,
A. Brid\v{z}ius$^{9}$,
D.W. Evans$^{15}$,
C. Fabricius$^{1,2,3}$,
M. Fiorucci$^{17}$,
U. Heiter$^{16}$,\newauthor
T.A. Kaempf$^{14}$,
A. Kazlauskas$^{8}$,
A. Ku\v{c}inskas$^{5,8}$,
V. Malyuto$^{13}$,
U. Munari$^{17}$,
C. Reyl\'e$^{18}$,\newauthor
J. Torra$^{1,2}$,
A. Vallenari$^{19}$,
K. Zdanavi\v{c}ius$^{8}$,
R. Korakitis$^{20}$,
O. Malkov$^{21}$,
A. Smette$^{22}$
\\
$^{1}$ Dept. Astronomia i Meteorologia, Universitat de Barcelona, Avda. Diagonal, 647, E-08028 Barcelona, Spain \\
$^{2}$ Institut d'Estudis Espacials de Catalunya (IEEC), Edif.Nexus, C/Gran Capit\`a, 2-4, 08034 Barcelona, Spain\\
$^{3}$ Niels Bohr Institute, Juliane Maries Vej 32, DK-2100, Copenhagen {\O}, Denmark \\
$^{4}$ Sterrewacht Leiden, P.O. Box 9513, 2300 RA Leiden, The Netherlands \\
$^{5}$ Lund Observatory, Lund University, Box 43, 221 00 Lund, Sweden\\
$^{6}$ Max-Planck-Institut f\"ur Astronomie, K\"onigstuhl 17, 69117 Heidelberg, Germany\\
$^{7}$ Observatoire de Gen\`eve, Chemin des Maillettes 51, CH-1290 Sauverny, Switzerland\\
$^{8}$ Institute of Theoretical Physics and Astronomy, Vilnius University, Go\v{s}tauto 12, Vilnius LT-01108, Lithuania  \\
$^{9}$ Institute of Physics, Savanoriu 231, LT-02300 Vilnius, Lithuania \\
$^{10}$ ESA/ESTEC, Research and Scientific Support Department, P.O.Box 299, 2200 AG Noordwijk, The Netherlands \\
$^{11}$ Institut d'Astrophysique et de G\'eophysique, Universit\'e de Li\`ege, All\'ee du 6 Ao\^ut 17, B-4000 Sart Tilman (Li\`ege), Belgium \\
$^{12}$ INAF - Osservatorio Astronomico di Torino, Strada Osservatorio 20, I-10025 Pino Torinese, Italy \\ 
$^{13}$ Tartu Observatory, 61602 T\~oravere, Estonia\\
$^{14}$ Sternwarte der Universit\"at Bonn, Auf dem H\"ugel 71, 53121 Bonn, Germany\\
$^{15}$ Institute of Astronomy, Madingley Road, Cambridge, CB3 0HA, UK \\
$^{16}$ Department of Astronomy and Space Physics, Uppsala University, Box 515, SE-75120 Uppsala, Sweden \\
$^{17}$ INAF - Osservatorio Astronomico di Padova, Sede di Asiago, 36012 Asiago (VI), Italy \\
$^{18}$ CNRS-UMR6091, Observatoire de Besan\c{c}on, BP 1615, F-25010 Besan\c{c}on, France \\
$^{19}$ INAF- Osservatorio di Padova, Vicolo Osservatorio 5, 35122 Padova, Italy \\
$^{20}$ Dionysos Satellite Observatory, National Technical University of Athens, Heroon Polytechniou 9, GR-15780 Zografos, Greece \\
$^{21}$ Institute of Astronomy, 48 Pyatnitskaya St. 119017 Moscow, Russia \\
$^{22}$ European Southern Observatory, Casilla 19001, Alonso de Cordova 3107, Vitacura, Santiago, Chile \\
}
\begin{document}

\date{}

\pagerange{\pageref{firstpage}--\pageref{lastpage}} \pubyear{2005}

\maketitle

\label{firstpage}

\begin{abstract}

The European {\Gaia} astrometry mission is due for launch in 2011. {\Gaia} will
rely on the proven principles of ESA's \textit{Hipparcos} mission to create an 
all-sky survey of about one billion stars throughout our Galaxy and beyond, by 
observing all objects down to 20th magnitude. Through its massive measurement of 
stellar distances, motions and multi-colour photometry it will provide
fundamental data necessary for unravelling the structure, formation and evolution 
of the Galaxy. This paper presents the design and performance of
the broad- and medium-band set of photometric filters adopted as the baseline
for {\Gaia}. The nineteen selected passbands (extending from the ultraviolet to
the far-red), the criteria, and the methodology on which this choice has been
based are discussed in detail. We analyse the photometric capabilities for
characterizing the luminosity, temperature, gravity and chemical composition of
stars. We also discuss the automatic determination of these physical parameters
for the large number of observations involved, for objects located throughout
the entire Hertzsprung-Russell diagram. Finally, the capability of the
photometric system to deal with the main {\Gaia} science case is outlined. 

\end{abstract}


%%%%%%%%%%%%%%%%%%%%  Introduction  %%%%%%%%%%%%%%%%%%%%%%%%%%%%%%%%%%%%

\section[]{Introduction}
\label{sec:introduction}

{\Gaia} has been approved as a cornerstone mission in the ESA scientific
programme. The main goal is to provide data to study the formation and
subsequent dynamical, chemical and star formation evolution of the Milky Way
galaxy \citep{perryman2001, mig}.  {\Gaia} will achieve this by providing an
all-sky astrometric and photometric survey complete to 20~mag in unfiltered
light. During the mission, on-board object detection will be employed and more
than 1~billion stars will be observed (as well as non-stellar objects to similar
completeness limits). The full mission (5-years) mean-sky parallax accuracies
are expected to be around 7~microarcsec at $V=10$, $12$--$25$~microarcsec at
$V=15$ and $100$--$300$~microarcsec at $V=20$ (depending on spectral type).
Multi-epoch, multi-colour photometry covering the optical wavelength range will
reach the same completeness limit.  Radial velocities
will be obtained for 100--150 million stars brighter than $V\simeq$17--18~mag
with accuracies of around 1--15~km s$^{-1}$, depending on the apparent
magnitude and spectral type of the stars and the sky density \citep[for details
see][]{RVSI,RVSII}.

The photometric measurements provide the basic diagnostics for classifying all
objects as stars, quasars, solar-system objects, or otherwise, and for
parametrizing them according to their nature. Stellar classification and
parametrization across the entire Hertzsprung-Russell diagram is required as
well as the identification of peculiar objects. This demands observation in a
wide wavelength range, extending from the ultraviolet to the far-red. The
photometric data must determine: (i) effective temperatures and reddening at
least for O-B-A stars (needed both as tracers of Galactic spiral arms and as
reddening probes), (ii) at least effective temperatures and abundances for
F-G-K-M giants and dwarfs, (iii) luminosities (gravities) for stars having large
relative parallax errors, (iv) indications of unresolved multiplicity and
peculiarity, and (v) a map of the interstellar extinction in the Galaxy. All of
this has to be done with an accuracy sufficient for stellar age determination in
order to allow for a quantitative description of the chemical and dynamical
evolution of the Galaxy over all galactocentric distances. Separate
determination of Fe- and $\alpha$-element abundances is essential for mapping
Galactic chemical evolution and understanding the formation of the Galaxy.

Photometry is also crucial to identify and characterize the set of
$\sim500\,000$ quasars that the mission will detect. Apart from being
astrophysically interesting in their own right, quasars are key-objects
for defining the fixed, non-rotating {\Gaia} Celestial {Re\-fe\-ren\-ce}
Frame, the optical equivalent of the International Celestial Reference
Frame \citep{mig}. On the other hand, {\Gaia} will identify about 900
quasars with multiple images produced by macrolensing.  Since this
number is sensitive to cosmological parameters, the {\Gaia} observations
will be able to constrain the latter.

Due to diffraction and the optical aberrations of the instrument, the
position of the centre of the stellar images is wavelength dependent. To
achieve the microarcsec accuracy level, astrometry has to be corrected for
this chromatic aberration through the knowledge of the spectral energy
distribution of the observed objects. Photometry is indispensable for
this. If uncorrected, chromatic errors could reach several milliarcsec,
cf.~Section~\ref{sec:chroma}.

As explained in the following sections, the photometric systems proposed during
the long development of {\Gaia} have been improved along with the increasing
collecting area of the telescopes, with better insight in the astrophysical
requirements, and with the development of mathematical tools to compare the
various proposed systems. The use of CCDs in a scanning astrometry satellite was
first proposed in 1992 \citep{hoeg93} as the \textit{ROEMER} project.  The
proposal included five broad passbands, \textit{UBVRI}, which would obtain much
better precision than the $B,~V$ of \textit{Hipparcos-Tycho} although with a
similar collecting aperture. The {\Gaia} collecting area has increased by up to
10~times with respect to \textit{ROEMER} and consequently the initial
photometric system has been upgraded several times. Eight medium-width passbands
were proposed by \citet{straihoeg95}, and spectrophotometry instead of filter
photometry was also considered in \citet{hoeg98}. A system of four broad and
eleven medium-width passbands was proposed by \citet{grenon} and adopted in the
`{\Gaia} Study Report' \citep{gsr}. The subsequent developments and updates have
yielded the present baseline with five broad- and fourteen medium-width
passbands.

This paper deals with the definition of the photometric system, the
relationship of its passbands with the stellar astrophysical diagnostics and the
evaluation of its performance in terms of the astrophysical parametrization of
single stars. This is based on the specific design implementation of the payload
commonly referred to as \Gaia-2
(cf.~Section~\ref{sec:instrument}) applicable at the end of the technology
assessment phase as of mid-2005.  The resulting astrometric and photometric
requirements form the basis of the industrial specifications for the satellite
implementation phase, with the consequence that the detailed design, due for
finalisation early in 2007, may differ in detail from the present description.
Nevertheless, the principles and objectives as well as the methods and
assessment tools described in this paper will remain applicable.

The paper is organized as follows: Section~\ref{sec:instrument} describes the
mission observation strategy, telescopes and focal planes. Section~\ref{sec:G}
deals with the measurement of the unfiltered light in the fields of view. The
principles of designing the multi-colour photometric system are outlined in
Section~\ref{sec:design_PS} and the purpose of the broad- and medium-passbands
is discussed in detail in Section~\ref{sec:BBPMBP}. Synthetic photometry and
corresponding error estimates are given in Section~\ref{sec:phot_precisions}.
The performance of the photometric system with respect to astrophysical
parameter determination is quantified in Section~\ref{sec:phot_performances}. In
Section~\ref{sec:galaxy} the potential of the photometric system for Galactic
structure and evolution studies as well as the performances for QSOs are
outlined. Finally, Section~\ref{sec:conclusions} and Appendix~\ref{ap:fom}
present the conclusions and describe the `Figure of Merit' mathematical tool for
the comparison of filter systems, respectively.

Throughout the rest of the paper we will use the following abbreviations:
AP: astrophysical parameter; 
BBP: broad-band photometer; 
C1B: set of passbands implemented in the Astro instrument; 
C1M: set of passbands implemented in the Spectro instrument; 
CCD: charge coupled device; 
FoM: figure of merit; 
MBP: medium-band photometer; 
MDM: minimum distance method;
NN: neural network; 
PS: photometric system;
QE: CCD quantum efficiency; 
QSO: quasar;
SED: spectral energy distribution;
SNR: signal-to-noise ratio;
ST: scientific target.

%%%%%%%%%%%%%%  Instrument description  %%%%%%%%%%%%%%%%%%%%%%%%%%%%%%%%%%%%

\section[]{Instrument description}
\label{sec:instrument}

 \begin{figure*}
  \begin{center}
    \leavevmode
 \epsfig{file=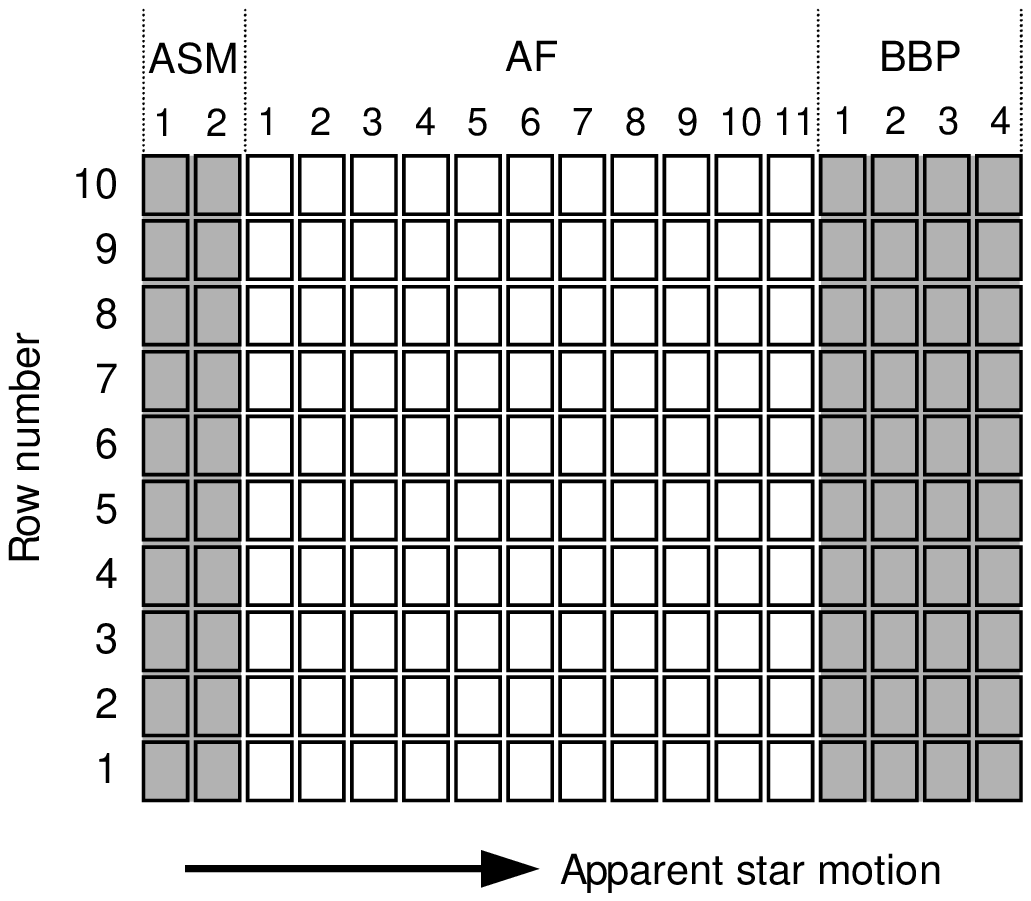,bbllx=192,bblly=243,bburx=500,bbury=494,width=0.85\columnwidth,silent,clip=}\epsfig{file=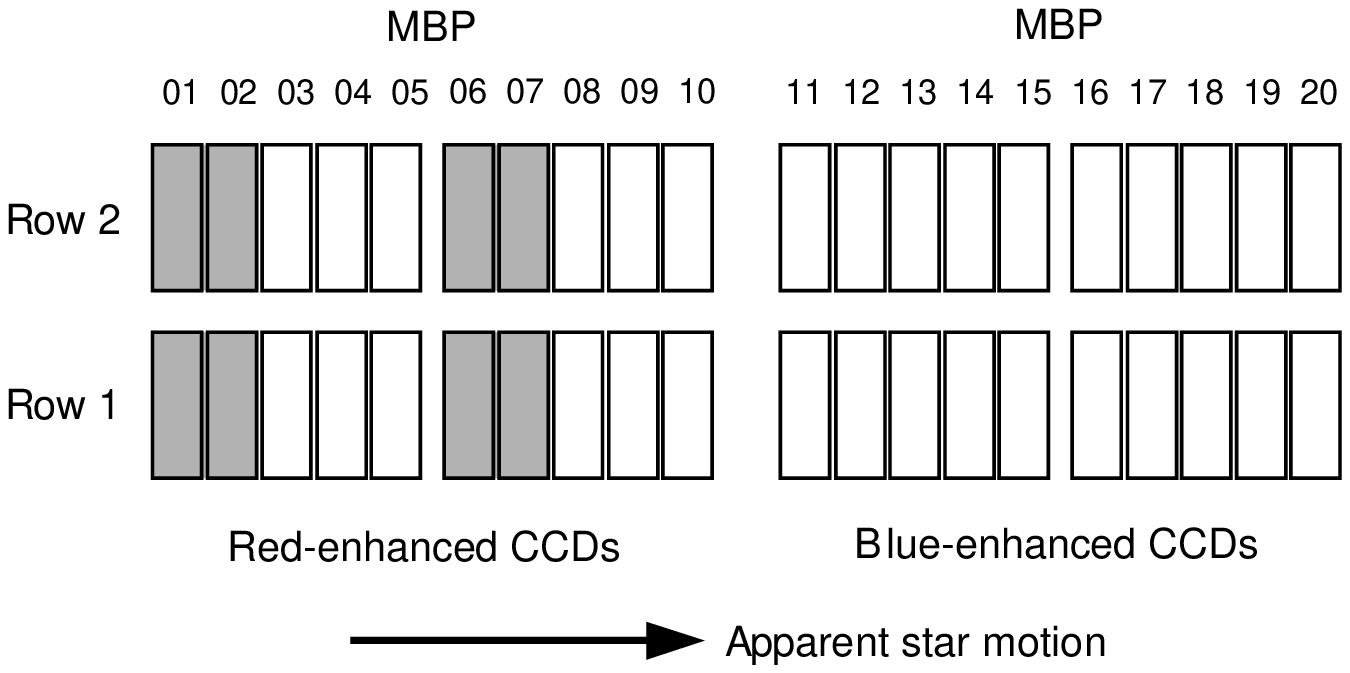,bbllx=192,bblly=300,bburx=580,bbury=494,width=1.1\columnwidth,silent,clip=}
% Explicit bounding box references are needed to avoid overlapping white space; different column widths are needed to obtain same-size text ("apparent star motion")
   \end{center}
  \caption{Schematic layout for the Astro field (left) and the Spectro/MBP field
  (right) in \Gaia-2. The designations BBP1--BBP4 and MBP01--MBP20 refer to the
  physical locations of the CCDs in the along-scan direction within the field,
  independent of their functionality (e.g., for detection or photometry) and the
  assignment of filters between them. The Spectro sky mapper, operating
  without filters, is physically located at MBP01, 02, 06, and 07.  ASM and AF
  are the Astro sky mapper and the Astro field, respectively.}
  \label{fig:FPAs}
\end{figure*}

\begin{table*}
  \caption{\Gaia-2 characteristics of the Astro and Spectro instruments.}
\label{tab:instruments}
\begin{center}
\begin{tabular}{llccc}
\hline
    &  Instrument                                             &Astro         &\multicolumn{2}{c}{Spectro\ \ \ \ \ \ \ \ }\\
    &  Photometer                                             &Broad-band    &Medium-band (blue) &Medium-band (red)\\
\hline
$F$ &Focal length [m]                                         &46.7          &2.3        &2.3\\
$A$ &Telescope pupil area [m$^2$]                        &1.4$\times0.5=0.7$ &0.5$\times0.5=0.25$ &0.5$\times0.5=0.25$\\
$n_{\rm sup}$ &Number of superimposed fields                  &2             &1          &1\\
$T_0(\lambda)$ &Telescope transmittance$^{(1)}$                 &$T^6_{\rm Ag}(\lambda)$ &$0.8 \times T^3_{\rm Al}(\lambda)$ &$T^3_{\rm Al}(\lambda)$\\
$\tau$ &Integration time [s] per single CCD                   &3.31          &12.0       &12.0\\
$Q(\lambda)$ &CCD quantum efficiency                          &CCD-green     &CCD-blue   &CCD-red\\
$r$ &Total detection noise [e$^-$~sample$^{-1}$]                &6.6           &7.3        &7.3\\
    &Available number of filter strips                         &4 (10 rows)   &10 (2 rows)&6 (2 rows)\\
    &Field of view across scan [deg]                          &0.7236        &1.4685     &1.4685 \\
$n_{\rm obs}$ &Mean number of observations per strip (5 years)$^{(2)}$ &83            &84         &84\\
    &Sample size on sky [arcsec $\times$ arcsec]$^{(3)}$              &0.0442$\times$1.5912  &0.897$\times$1.346 &0.897$\times$1.346\\
    &Double-star resolution [arcsec]                          &0.05--0.1     &0.5--1     &0.5--1\\
&No-filter magnitude                                          &$G$           &--         &$GS$ \\
\hline
\multicolumn{5}{l}{\footnotesize{$^{(1)}$ $T^n_{\rm X}(\lambda)$ denotes $n$ reflections of material X. The factor 0.8 in MBP-blue comes from the dichroic beam splitter.}} \\
\multicolumn{5}{l}{\footnotesize{$^{(2)}$ The number of observations with Astro ranges from $\sim$40 to 250 and in Spectro from $\sim$50 to 220.}} \\
\multicolumn{5}{l}{\footnotesize{$^{(3)}$ Sample size on the sky: A sample is a binning of CCD pixels in an area along-scan $\times$ across-scan.}}
\end{tabular}
\end{center}
\end{table*}

\subsection{Observation strategy}

{\Gaia} is a survey mission. Operating from a Lissajous orbit around the second
Lagrange point of the Sun-Earth/Moon system (L2), the satellite will
continuously observe the sky. During its 5-year mission, {\Gaia} will rotate, at
a fixed speed of 60~arcsec s$^{-1}$, around a slowly precessing spin axis. As a
result of this spin motion, objects traverse the focal planes, which have
viewing directions oriented at right angles to the spin axis. The along-scan
direction is defined to be the direction in which the images move in the
focal plane, and the across-scan direction as the perpendicular direction.
Thus, an object crosses the field of view in the along-scan direction at a 
roughly constant across-scan coordinate.

\Gaia's observations are made with high-quality, large-format charge coupled
devices (CCDs). These detectors are operated in time delay and integration
mode, with charge images being transported (clocked) in synchrony with optical
images moving across the field due to the rotation of the satellite.
Each CCD may, in principle, be equipped with a filter (an interference, standard
coloured glass or multi-layer filter) defining, together with the telescope
transmission and the CCD quantum efficiency (QE), a certain photometric
passband.  By design, the arrangement of CCDs and filters in the focal planes
ensures redundancy in case of failure.

Stars brighter than $\sim$12~mag pose a special challenge. Pixel saturation may
be avoided for such objects by the activation of gates on the CCD, effectively
reducing the CCD integration time.

\subsection{{\Gaia} instruments}

{\Gaia} will perform measurements by means of two physically distinct
instruments, with different viewing directions: the Astro instrument, designed
for astrometric and broad-band photometric observations, and the Spectro
instrument, used for medium-band photometry and radial-velocity measurements.
Astro and Spectro differ in spatial resolution, available integration time, the
number and type of passbands that can be used, and in telescope transmission and
detector characteristics (Table~\ref{tab:instruments} and
Figure~\ref{fig:FPAs}).

The astrometric focal plane incorporates three functions: (i) the sky mapper;
(ii) the main astrometric field; and (iii) the broad-band photometer
(BBP; Section~\ref{sec:section_BBP}). BBP is mainly aimed at sampling the
spectral energy distribution of objects over a wide wavelength range to allow
on-ground correction of image centroids measured in the main astrometric field
for systematic chromatic shifts caused by aberrations. In addition, BBP
measurements contribute to the astrophysical characterisation of objects,
especially in dense stellar fields.

Through implementation of a dichroic beam splitter, the Spectro instrument
serves two distinct focal planes: one for the radial-velocity spectrometer
(RVS), and one for the medium-band photometer (MBP;
Section~\ref{sec:section_MBP}).  The Spectro/MBP focal plane incorporates two
functions: (i) a sky mapper; and (ii) the MBP instrument. The main goal of
MBP is to determine the astrophysical parameters of objects, which, in
combination with the astrometric measurements, will enable astronomers to
fulfil \Gaia's main science objective. 
RVS observes a spectral region around 860~nm at a nominal resolution of
$11\,500$.  Its main aim is to determine radial velocities for bright stars,
down to $V\simeq17$--$18$~mag.
For the brightest ones, astrophysical parametrization is also foreseen.

The angular resolution of the instruments allows photometry to be obtained at
stellar densities up to $750\,000$ stars/sq~deg in the BBP and up to
$200\,000$--$400\,000$ stars/sq~deg in the MBP (cf.\ Section~\ref{sec:crowded}).
The resolution limit for double stars is about $0.05$--$0.1$ arcsec in the Astro
field and $0.5$--$1$ arcsec in the MBP.

\subsection{Sky mappers}
\label{sec:skymappers}

Both sky mappers work in unfiltered light and allow autonomous object
detection and confirmation, including rejection of prompt particle events
(cosmic rays, solar protons, etc.). Detection and confirmation
probabilities are a function of magnitude and object density in the field
of view. They are effectively unity up to the survey limit (20~mag)
dropping quickly to zero for fainter objects \citep{arenou}. On-board
object detection has several advantages: (i) sampling of detected objects
can be limited to `windows', i.e., areas centred on the object
\citep{hoeg}; this allows to flush useless pixels containing empty sky,
which is of benefit to both CCD read-out noise and the data volume to be
transmitted to ground; (ii) it allows the unbiased detection of all
objects, assuring that the resulting catalogue will be complete to the
survey limit; and (iii) unpredictable `peculiar objects', such as
supernovae or solar-system objects, will be observed, if brighter than the
detection limit. Moreover, on-board detection is mandatory as no input
catalogue exists which is complete to the survey limit at the {\Gaia}
spatial resolution.

\subsection{Astro/BBP}

\Gaia's astrometric observations are made without a filter in order to
minimize photon noise. The mirror coatings and CCD QE effectively define a
broad (white-light) passband, called $G$ (Section~\ref{sec:G}). The Astro
focal plane receives the light from the superposition of two viewing
directions on the sky, separated by the so-called `basic angle', of order
$100^\circ$.  This superposition of the two fields of view will lead to a
number of complications in the data processing, but throughout this paper
we only consider the first order effect of the doubling of the diffuse
sky-background.

The BBP consists of 40 CCDs, arranged in 4 across-scan strips and 10 along-scan
rows (the {\Gaia} naming convention for the along-scan direction is rows of CCDs
and columns of CCD pixels, and across-scan we speak of strips of CCDs and lines
of pixels). In principle, the filters can be distributed among the 40 CCDs in
any combination that is desirable from a scientific point of view.  A preferable
situation, for example for variable-star science, would be to have identical
filters on each of the 10 CCD rows within a single CCD strip: this would yield
quasi-simultaneous measurements in the different passbands on each field-of-view
crossing, independent of an object's across-scan coordinate. Such a solution
would, however, limit the number of BBP passbands to the number of CCD strips,
i.e., 4.

\subsection{Spectro/MBP}

The MBP consists of 40 CCDs arranged in 20 strips and 2 rows. The CCDs in the first
10 strips are illuminated directly from the Spectro telescope. These CCDs are
red-enhanced, which means they are thicker than `normal CCDs' and have a
red-optimized anti-reflection coating. The CCDs in the last 10 strips receive
only the blue light from the Spectro telescope.  These detectors are
blue-enhanced, which means they have the same thickness as `normal CCDs' but a
blue-optimized anti-reflection coating. Four of the red-enhanced strips act as
sky mappers and do not have filters; the associated broad passband is called
$GS$. The remaining 6 strips with red- and the 10 strips with blue-enhanced CCDs
may be equipped with filters, defining up to 16 different passbands, assuming
the two CCD rows in each strip have identical filters (for the sake of
redundancy). One of the red passbands will cover the same spectral region as the
RVS, around 860 nm, leaving 15 free strips.

%%%%%%%%%%%%%%   G and G_S  %%%%%%%%%%%%%%%%%%%%%%%%%%%%%%%%%%%%%%%%%%%%%%

\section[]{The $G$ and $GS$ passbands}
\label{sec:G}

\begin{figure}
  \begin{center}
    \leavevmode
 \epsfig{file=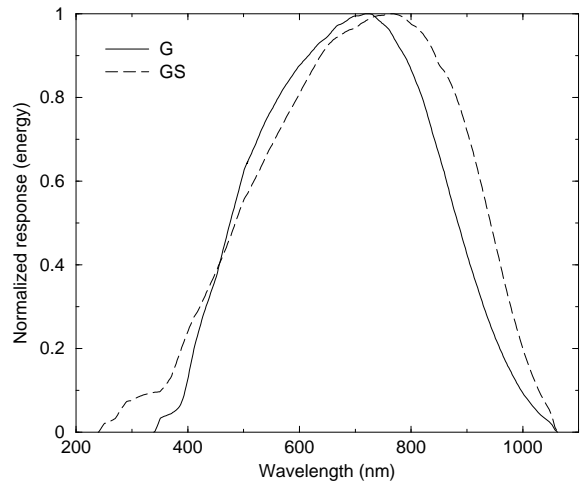,width=0.9\linewidth}
   \end{center}
  \caption{The $G$ and $GS$ {\Gaia} broad passbands corresponding to the white light
  observations in the Astro and Spectro focal planes, respectively.}
  \label{fig:QE}
\end{figure}

\begin{figure}
  \begin{center}
    \leavevmode
 \epsfig{file=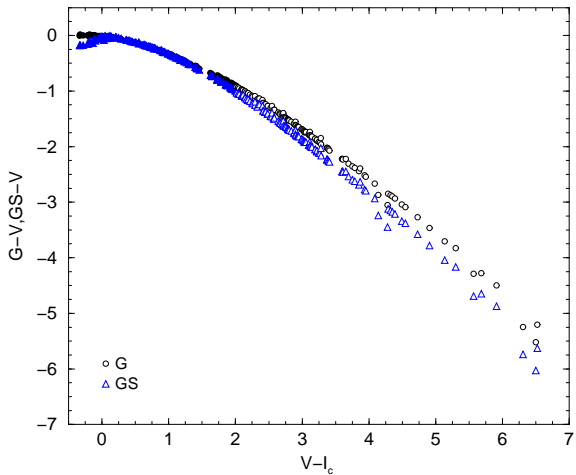,width=0.9\linewidth}
   \end{center}
  \caption{The relation between $G-V$ and $GS-V$ and $V-I_C$ for the white light
  passbands in the Astro and Spectro focal planes.  Every star in the spectral
  library of \citet{pickles} is represented by a filled symbol. Open symbols
  correspond to the same stars reddened by \av$=5$ mag. Reddening vectors run
  parallel to the colour-colour relationship. Pickles' library was chosen because
  it extends to very red objects. No appreciable differences for stars with
  \teff$\ge 4000$~K are obtained when using spectral energy distributions from
  the BaSeL2.2 library or other libraries.}
  \label{fig:GV}
\end{figure}

The $G$ and $GS$ passbands corresponding to the white light observations in the
Astro and Spectro instruments, respectively, are shown in Fig.~\ref{fig:QE}.
They cover the wavelength range from 400 to 1000~nm and 350 to 1025~nm, with the
maximum energy transmission at $\sim715$ and $\sim765$~nm and the full width at
half maximum of 408 and 456~nm, respectively for $G$ and $GS$. The relation
between the associated magnitudes and the Johnson $V$ magnitude is shown in
Fig.~\ref{fig:GV}.  Very red objects (either intrinsically red or highly
reddened) are much brighter in $G$ and $GS$ than in $V$, and therefore, the
{\Gaia} limiting magnitude of $G_\mathrm{lim}\sim20$ translates into
$V_\mathrm{lim}\sim 20$--$25$, depending on the colour of the observed object.

\begin{figure}
  \begin{center}
    \leavevmode
 \epsfig{file=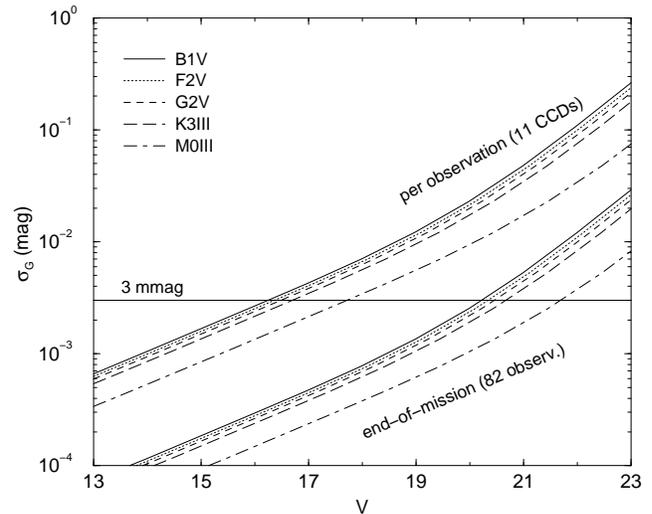,width=1.0\linewidth}
   \end{center}
  \caption{Estimated precision of the $G$ magnitude per focal plane transit and
  at the end of the mission according to equation (\ref{eq:eqerror}) assuming
  $\sigma_{\rm cal}=0$. In this paper, the contribution of calibration error to
  the end-of-mission precision is assumed to be 3~millimag (horizontal line).}
  \label{fig:errorsG}
\end{figure}

The estimated precisions for the $G$-magnitudes per focal plane transit and at
the end of the mission, computed as described in Section~\ref{sec:aperture}, are
shown in Fig.~\ref{fig:errorsG}. Taking into account the photon noise from the
source, the background and the read-out noise, precisions of $\sim 10$ and $\sim
1$~millimag are achievable at $V\sim 19$. This implies that the precision of the
$G$ measurements is ultimately limited by the calibration errors (see
Section~\ref{sec:calibration}).

{\Gaia} will provide distances at the 10\% accuracy level for some 100--200
million stars, which combined with estimates of the $G$ magnitude and the
interstellar extinction, will yield unprecedented absolute magnitudes, in both
accuracy and number.

The $G$ passband also yields the best signal to noise ratio for
variability detection among all the {\Gaia} passbands.
{\Gaia} will monitor millions of variable stars (eclipsing binaries,
Cepheids, RR~Lyrae, Mira-LPVs, etc).  \citet{eyer} and \citet{eyermignard}
provide comparisons with other variability surveys and a detailed
discussion of the effects of the variable time sampling and number of
observations due to the scanning law.

%%%%%%%%%%%%%%  Designing PS %%%%%%%%%%%%%%%%%%%%%%%%%%%%%%%%%%%%%%%%%%%%%%

\section[]{Designing the photometric system}
\label{sec:design_PS}

Many ground-based photometric systems (PSs) exist but none satisfies all the
requirements of a space-based mission such as \Gaia: portions of the spectrum
limited on-ground by telluric O$_3$ opacity in the blue and O$_2$ and H$_2$O
absorption bands in the red are accessible to \Gaia; existing PSs have usually
been designed for specific spectral type intervals or specific objects, while
{\Gaia} photometry must deal with the entire Hertzsprung-Russell diagram, must
allow taxonomy classification of solar-system objects and must be able to 
identify quasars and galaxies. In
addition, {\Gaia} allows the extension of stellar photometry to Galactic areas
where the classical classification schemes may be no longer fully valid because
of systematic variations in element abundances in stellar atmospheres and in
interstellar matter. Finally, {\Gaia} has to astrophysically characterize
objects over a very large range of brightness, from $G\sim6$ to the faint limit
of $G_{\rm lim}\sim 20$, and consequently the width of the passbands will
reflect a trade-off between sensitivity to physical parameters and the
possibility to measure faint stars. Therefore, designing a new PS is the best
approach to ensuring that \textit{Gaia}'s ambitious goals will be achieved.
This became clear early on when initial efforts to use existing PSs showed that
these failed to cover all {\Gaia} requirements. A fairly complete census of
existing PSs can be found in \citet{vytas1992,moro2000,fiorucci,bessell2005}.

Criteria for the design of a new PS have to be established \textit{a priori}. As
it is widely known, narrow passbands are very efficient in measuring specific
spectral features, but have low performance for faint objects; broad passbands
yield low photon-noise for faint objects but cannot give one-to-one
determination of astrophysical parameters. The ultraviolet contains
important information on astrophysical parameters, but {\Gaia} will not be
able to measure faint red objects in this spectral range.
Therefore, a compromise between the different options (number, location,
and width of the passbands and their total exposure time) is needed
to achieve a PS that is maximally capable of separating objects with 
different astrophysical parameters in filter flux space.
Such a PS should allow both accurate \textit{discrete source classification}
(i.e.\ identification of stars, galaxies, QSOs, solar-system objects, etc.) and
\textit{continuous parametrization} (values of astrophysical parameters, e.g.\
for stars {\teff}, {\feh}, etc) for all {\Gaia} targets across the astrophysical
parameter space.
 
\subsection{Principles for the design}
\label{sec:principles_PS}

Over the years of mission study and development, the instrumental concept of
{\Gaia} has evolved.  As a consequence, the requirements and constraints for
designing the PS have evolved in terms of the number of passbands, the exposure
time available per passband, the wavelength coverage in BBP and MBP, the spatial
resolution of Astro and Spectro, the goals of BBP and MBP, the complementarity
of BBP, MBP and RVS measurements, etc.

With the consolidation of the study-phase of the {\Gaia} instrument design, the
requirements for BBP and MBP are stabilized. The PS accepted as the baseline for
the mission and presented in this paper is based on the following constraints:
\begin{itemize}
\item The instrument design is as described in Section~\ref{sec:instrument}.
\item Photometry from Astro has to account for chromaticity in order to achieve
  microarcsec astrometry, which implies a measurement of the spectral energy
  distribution of each object with contiguous broad passbands covering the whole
  wavelength range of the $G$ passband (see Section~\ref{sec:chroma}).
\item The photometry from Spectro has to provide the astrophysical characterization
  of the observed objects.
\item The photometry from Astro has to provide the astrophysical
  characterization of the observed objects in dense fields (at stellar densities
  larger than $200\,000$--$400\,000$ stars/sq~deg; mainly in the bulge, some
  Galactic disk areas and globular clusters).
\item One of the medium passbands in Spectro has to measure the flux in the
  wavelength range covered by RVS (848--874~nm).
\item The trigonometric parallax is available and has to be used to constrain
  the luminosity (or the absolute magnitude).
\item The information from RVS is not used for designing the PS because of the
  different limiting magnitude and spatial resolution of that instrument (in
  the actual {\Gaia} data processing it is foreseen that astrometry, photometry
  and RVS measurements will be used together to derive astrophysical parameters,
  when possible).
\item The {\Gaia} PS is optimized for single stars, where priority is given to
  those types crucial for achieving the {\Gaia} core science case, namely the
  unravelling of the structure and the formation history of the Milky Way. These
  stars are named `scientific photometric targets' or simply `scientific
  targets' (ST; see Section~\ref{sec:test_pop}) and they constitute our test
  population.
\item Every ST is characterized by its astrophysical parameters (APs), where we
  consider \teff, \logg, chemical composition and \av\footnote{Throughout the
  paper {\av} means $A_{\lambda=550}$ i.e.\ the monochromatic interstellar
  extinction at $\lambda=550$~nm}. The chemical composition is described by the
  metallicity, {\feh}, and the $\alpha$-element enhancements, {\afeh}.
\item The error goals for the AP determination are established for every ST
  (\skgoal; see Section~\ref{sec:fom} and Appendix~\ref{ap:fom}).
\item The actual performance of a given PS with respect to the error goals is
  measured using an objective `Figure of Merit' (see Section~\ref{sec:fom} and
  Appendix~\ref{ap:fom}).
\item The global degeneracies of the PS have to be evaluated (see
  Section~\ref{sec:degen}).
\item Additional merits such as, for example, the performance with respect to
  discrete object classification, the performance for non-ST objects, etc.  have
  to be considered.
\end{itemize}

The procedure to come to a baseline PS \citep{brown} is based on the
maximization of this `Figure of Merit', a minimization of the global
degeneracies and an evaluation of additional merits.

\begin{table*}
  \caption{The main science goals for {\Gaia} concerning Galactic structure and
  evolution studies and the main corresponding tracers, from the `{\Gaia}
  Concept and Technology Study Report' \citep{gsr}.}
\label{tab:tracers}
\begin{center}
\begin{tabular}{ll}
\hline
Science goal& Main tracers \\
\hline
Chemical abundance galactocentric distribution& G and K giants \\
Galactocentric age gradients, star formation rate & HB stars, early-AGB, main sequence turn-off stars \\
Disentangling age-metallicity degeneracy& main sequence turn-off, sub-giants \\
Star formation history& main sequence stars earlier than G5 V, sub-giants \\
Detailed knowledge of luminosity function& low main sequence stars \\
Halo streams, age and chemical abundance determinations& G and K giants \\
Outer halo ($R>20$ kpc), accretion and merging& G and K giants and HB stars \\
Earliest phases of evolution of the Galaxy& most metal-poor stars,
C- sub-giants and dwarfs near the turn-off \\
Distance scale& RR-Lyrae, Cepheids \\
Thick disk formation mechanism (pre- vs.\ post-thin disk)& G and K giants, HB stars \\
Merging and/or diffusion& high velocity A-type stars \\
`In situ' gravitational potential, $K_z$, age-velocity relation& F-G-K dwarfs \\
Large scale structure (warp, asymmetry)& K-M giants\\
Interstellar medium distribution& O-F dwarfs  \\
Large scale structure (spiral arms, star-formation regions)& OB stars, supergiants \\
Star formation rate in the bulge& red giant branch, asymptotic giant branch \\
Shape of the bulge, orientation, bar& red giant branch, asymptotic giant branch, Red clump stars \\
\hline
\end{tabular}
\end{center}
\end{table*}
                                                                                                  
\subsection{Figure of Merit}
\label{sec:fom}

For a given photometric system the `Figure of Merit' (FoM) is constructed by
calculating for each ST and each of its APs, $p_k$, the ratio $\skpost/\skgoal$.
Here {\skpost} is the estimate of the error than can be achieved with the given
PS for $p_k$ and {\skgoal} is the above mentioned error goal. The procedure for
estimating {\skpost} and the definition of the FoM were proposed by
\citet{lin_fom}.  For a given AP, $p_k$, and a PS with measured fluxes $\phi_j$
($j=1,n$ passbands), {\skpost} is estimated using the sensitivity of the PS to
that parameter ($\partial\phi_{j}/\partial p_k$) and the errors of the
photometric observations ($\epsilon_j$). The latter are based on the noise model
for the instrument-PS combination (see Section~\ref{sec:aperture}). The details
are given in Appendix~\ref{ap:fom}.

The global FoM as given in equation (\ref{eq:calcq}) in Appendix~\ref{ap:fom} is a
weighted sum of the individual FoMs of every ST, which in turn are weighted sums
of the ratios $\skpost/\skgoal$ for each of the APs $p_k$ (equation \ref{eq:calcqi}).
A higher value of this FoM indicates a better performance for the given PS. The
global FoM thus describes the performance of a PS across the HR-diagram by
taking into account the errors that can be achieved and the relative priorities
of the different STs and their APs. The local degeneracies in the AP
determinations (i.e., correlations between the errors for different APs for a
given ST, see Section~\ref{sec:degen}) are also taken into account in the FoM
formalism.  Each local degeneracy between APs leads to increased standard errors
{\skpost} and thus a lower FoM.

In our implementation the derivatives $\partial\phi_{j}/\partial p_k$ and the
error estimates {\skpost} are calculated numerically from simulated photometric
data (Section~\ref{sec:simulation}). The calculation thereof requires
(synthetic) spectral energy distributions (SEDs) of the STs and a noise model
for the photometric instruments. The matrix \mat{B} from equation (\ref{eq:cpost})
includes the {\it a priori} information for the AP vector \vect{p} and in our
case corresponds simply to the range of possible values of the parameters $p_k$.
The information from the parallax and its error is incorporated into \mat{B}
following \citet{lin_para}.  Other information could be added (known reddening
in a certain Galactic locations, ranges of abundances according to Galactic
population, etc) but this will introduce our preconceptions of the structure and
stellar populations of the Galaxy into the PS design and we therefore did not
include such constraints.

The MBP data will have a lower spatial resolution than the BBP data. For
non-crowded regions on the sky (with respect to MBP) we assume that the
MBP and BBP photometry will always be combined for the estimation of
APs. Hence the calculation of the achievable
posterior errors is always done by combining BBP and MBP data. For dense
stellar fields, only BBP will be available and in that case the achievable
AP errors and the FoM have been computed using only broad-band photometry.

Spectral energy distributions of the STs in the test population were taken
from the BaSeL2.2 \citep{lejeune}, NextGen \citep{haus1} and
MARCS\footnote{see also http://marcs.astro.uu.se} \citep{gustafsson}
libraries. The BaSeL2.2 library is a compilation of synthetic spectra from
libraries published by \citet{kurucz}, \citet{bessell}, \citet{fluks} and
\citet{haus1}.  It covers the whole HR-diagram and {\feh} abundances from
$-5$ to $+1$~dex, but with solar $\alpha$-element abundances.  A new
version of the NextGen library (NextGen2) has been built taking into
account {\Gaia} mission needs \citep{haus2}. This library includes SEDs
for stars cooler than $10\,000$~K with {\feh} ranging from $-2$ to
$+0$~dex and {\afeh} from $-0.2$ to $+0.8$~dex. A more extended grid is
currently available \citep{brott}. Finally, a new version of the MARCS
library taking into account non-solar $\alpha$-element abundances has been
created specifically for {\Gaia} studies. This library provides coverage
between 3000~K and 5000~K, {\feh} from $-4$ to $+0.5$~dex and {\afeh} from
$+0$ to $+0.4$~dex.

Empirical libraries, like those by \citet{pickles} and \citet{gunn}, do
not provide full coverage of the HR-diagram and the corresponding chemical
composition range, and therefore they are not appropriate for computing
the FoM values.

Finally we would like to make a few remarks here about the interpretation of the
calculated values of {\skpost} and the corresponding FoM for a set of proposed
PSs. The values of {\skpost} calculated as described in Appendix~\ref{ap:fom}
should not be taken as the actual errors that will be achieved by the {\Gaia}
photometric system. They represent the achievable precision if the synthetic
spectra represent the true stars and if the noise model is correct.  This issue
is discussed more thoroughly in Section~\ref{sec:phot_performances}.  What makes
the FoM a powerful tool is that it enables an objective \textit{comparison} of
different PS proposals, based on a set of agreed error goals and scientific
priorities. In addition for each PS a detailed study can be made of its
strengths and weaknesses as compared to other PSs by examining the FoM for
individual STs and for groups of STs (such as specific types of stars,
populations in certain Galactic directions, bright versus faint stars, reddened
versus unreddened, etc). Once the best PS proposal has been chosen it can be
further tuned by using the FoM procedure to improve its sensitivity to certain
APs or to improve the performance for certain groups of stars. This objective
figure of merit approach has not been used before in the design of a photometric
system.

\subsection {Test population and error goals}
\label{sec:test_pop}

According to the scientific goals of {\Gaia} (see
Table~\ref{tab:tracers}), for every Galactic stellar population several
kinds of stars were selected as STs and a priority, expressed as a
numerical weight, was assigned to them \citep{jorditargets,jordiprio}.
These stars have been considered in different directions in the Milky Way
(toward the centre, the anti-centre and perpendicular to the Galactic
plane) and at different distances ($0.5$, 1, 2, 5, 10 and 30~kpc) in
accordance with a Galaxy model \citep{figueras}. The STs have been
reddened according to a 3D interstellar extinction model \citep{drimmel},
assuming a standard extinction law ($R_V=3.1$). Bulge stars in areas of
high and low interstellar extinction were also included. This results in a set
of about 6500 targets each with a parallax error estimated from its colour and
apparent magnitude \citep{gsr}. The targets in the halo are considered to have
{\feh} abundances from $-4$ to $-1$~dex and {\afeh} abundances from $+0.2$ to
$+0.4$~dex; for the thick disk targets {\feh} ranges from $-2$ to $0$~dex and
{\afeh} from $0.0$ to $+0.2$~dex; for the thin disk targets {\feh} ranges from
$-1$ to $+0.5$~dex and {\afeh} from $-0.2$ to $0.0$~dex; and finally for the
bulge targets we assume the same range for {\feh} as for the thin disk and
{\afeh} ranges from 0 to 0.4~dex. The ranges of temperatures and absolute
luminosities are assigned according to the specific range of ages of each
stellar population. The values of $T_{\rm eff}$ range from $3000$ to $40\,000$~K
and the values of {\logg} are in the range $0.0$--$5.0$~dex.

We adopted the following precision goals (\skgoal):
\begin{itemize}
\item $T_{\rm eff}$ for A-M stars: $\sigma_{T_{\rm eff}}/T_{\rm eff} =1-2$\%
\item $T_{\rm eff}$ for O-B stars: $\sigma_{T_{\rm eff}}/T_{\rm eff} =2-5$\%
\item \av: $\sigma_{\av}=0.1$~mag  at $\av\le3.0$ mag, $\sigma_{\av}=0.5$~mag
at $\av>3.0$ mag
\item {\mv} for stars with $\sigma_{\pi}/\pi\le10$\%: assumed known
\item {\logg} for stars with $\sigma_\pi/\pi>10$\%: $\sigma_{\logg}=0.2$~dex
\item {\feh} (not to be determined for OB stars and supergiants): $\sigma_{\feh}=0.1$~dex
\item {\afeh}: $\sigma_{\afeh} < 0.3$~dex
\item AGB: Carbon/Oxygen classification
\end{itemize}

\subsection{Photometric system proposals}
\label{sec:proposals}

In this paper we discuss in detail only the baseline PS for {\Gaia} that results
from the optimization process described above. However, many proposals for a PS
for {\Gaia} have been studied.

Different approaches have been used for the definition of the passbands.
\citet{munari}, \citet{grenon}, \citet{vytas2000}, \citet{jordi2F, jordi3F},
\citet{vladas1X, vladas} and \citet{lin2003a} based their BBP and/or MBP
proposals on their astrophysical expertise and/or the performance of the
existing photometric systems.  \citet{taut2003} and \citet{taut2005} provided
guidelines for designing a PS sensitive to C,N,O and $\alpha$-process elements.
We note that the sets of filters by \citet{munari}, \citet{grenon}, and
\citet{vytas2000} were designed with substantially different constraints imposed
in early phases of the {\Gaia} instrument design.

\citet{coryn} developed a novel method for designing PSs via a direct
numerical optimization. By considering a filter system as a set of free
parameters (central wavelengths, FWHM etc.), it may be designed by
optimizing a figure of merit with respect to these parameters. This figure
of merit is a measure of how well the filter system separates the stars in
the data space and of how much it avoids degeneracies in the AP
determinations. The resulting filter systems tend to have rather broad and
overlapping passbands and large gaps in between at the same time. Some
commonalities with conventional PSs may be recognized. Although first
analyses showed these systems to yield FoM values only slightly inferior
to conventional PS proposals, the method was considered too novel and
further studies were not pursued. 

\citet{lin2001} proposed several BBP systems, differing in the number of
passbands and their profiles, and tested their performance with respect to
chromatic effects estimation. He concluded that five passbands do not provide a
clear advantage over four passbands, and that the overlapping of the passbands
is more important.

\citet{heiter} used the Principal Components Analysis technique to design a set
of BBP passbands which is optimal for a subset of STs with effective
temperatures between 3000 and 5000~K. Photometry was simulated for a grid of
sets of four broad filters each.  The FWHMs were varied in equidistant steps,
while the central wavelengths were determined by requiring contiguous filters
covering the whole wavelength range.
The resulting set of passbands was found to perform better than all other
proposals for bulge stars but worse than all others when evaluated for the
complete set of STs \citep{jordievalBBP}.

All proposed photometric systems were evaluated using the FoM procedure
and the results were returned to the proposers who were then given the
opportunity to refine their respective PSs and submit
improved versions as a new PS proposal. Updates of the PSs above or new
ones came out of every evaluation cycle (\citealt{vytas2004},
\citealt{knude}, \citealt{hoegknudea,hoegknudeb} and \citealt{jordibbp,
jordimbp}). Finally, after several trials for the fine tuning of a few
individual passbands, the C1B and C1M sets 
\citep{jordievalBBP,jordievalMBP} were adopted as the baseline PS for
{\Gaia}.  Detailed results of the evaluation of all PS proposals can be
found in \citet{jordievalBBP,jordievalMBP}.

\subsection {Local and global degeneracies}
\label{sec:degen}

The goal of designing a photometric system that allows both accurate discrete
classification and continuous parametrization of observed sources translates to
demanding that the degeneracies in the PS are minimized.

The discrete classification problem concerns the way very different parts of AP
space are mapped by the PS into the filter flux space. For example, due to
extinction a highly reddened O-star may to first order look like a nearby
unreddened cool dwarf. A good PS should as much as possible be free of such
`global' degeneracies. The continuous parametrization problem concerns the way a
PS maps a small region around a certain object in AP space onto filter fluxes.
Here it is the `local' degeneracies that are important. Examples include the
well know degeneracy between {\teff} and {\av} and the difficulty of
disentangling the effects of {\teff} and {\logg} for a PS without passbands
shortward of the Balmer jump.

Our method of evaluating the proposed PSs as outlined in Section~\ref{sec:fom}
focuses on the local degeneracies. Consider the gradient vectors which describe
how the filter fluxes respond to changes in particular APs. For a good PS these
gradients should be large with respect to the noise in the data and they should
ideally be orthogonal to each other, where the orthogonality is also defined
with respect to the noise (i.e., for the gradients with components
$1/\epsilon_j\times\partial\phi_j/\partial p_k$, see Appendix~\ref{ap:fom}).
Large gradients mean that the PS is sensitive to the corresponding APs while
their orthogonality ensures that there are no local degeneracies. The FoM takes
this into account by calculating the posterior errors on estimated APs using the
sensitivity matrix which contains these gradient vectors (see
Appendix~\ref{ap:fom} for more details). Small gradient vectors are reflected in
larger errors on the estimated AP. Non-orthogonal gradient vectors will also
lead to larger errors and to non-zero covariances (i.e., correlated
errors) in the posterior variance-covariance matrix of the estimated APs. Both
effects will lead to a lower FoM (this is explained in more detail in
Appendix~\ref{ap:fom}).

The FoM calculations do not take global degeneracies into account. In fact
it is assumed that one already has available a good classification of the
object to be parametrized so that the linearized equations from which the
FoM is derived apply. The global degeneracies reflect the highly
non-linear mapping from AP space to filter flux space and are difficult to
characterize in practice.

We attempted to compare the different PS proposals with respect to global
degeneracies by employing self-organizing maps to explore how the
different STs cluster in data space and how well they can be separated.
The results were inconclusive as the different PS proposals showed rather
similar behaviour. This is plausible because to first order all the
proposed filter systems were similar. The sets of BBP passbands were very
much alike and the sets of MBP passbands all had a set of blue and a set
of red filters with a gap between these sets from $\sim$550 to
$\sim$700~nm and they all had an H$\alpha$ filter in this gap. With a
continuous sampling of stellar parameters a filter system will define a
complex manifold in the space of filter flux vectors onto which each star
will be mapped. It is the overall shape of this manifold that determines
the presence or absence of global degeneracies in the PS.
Hence it may be that the filter systems that had been considered all
define roughly the same manifold, the differences between filter systems
only being manifest at the local level (where the FoM calculations are
more relevant).

Characterizing the global degeneracies will be an important task in the
context of the automatic classification effort for the {\Gaia} data
processing. A full understanding of the behaviour of the {\Gaia} PS is
essential for setting up appropriate discrete classification and
continuous parametrization algorithms.

%%%%%%%%%%%%%%  BBP & MBP  %%%%%%%%%%%%%%%%%%%%%%%%%%%%%%%%%%%%%%%%%%%%%%

\section{The {\Gaia} photometric system}
\label{sec:BBPMBP}

In this section we describe in detail the baseline photometric system for
{\Gaia}, called C1, which consists of the C1B and C1M broad and medium
passbands. The role of each of the passbands is described in relation to
spectral features and astrophysical diagnostics.

\subsection{The C1B broad passbands}
\label{sec:section_BBP}

\begin{figure}
\begin{center}
\leavevmode
\epsfig{file=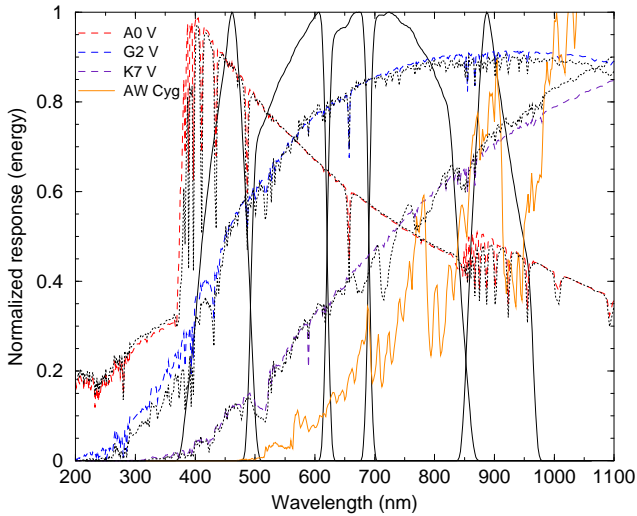,width=1.0\linewidth}
\end{center}
\caption{Response curves of the filters in C1B folded with the optics
transmission and the QE of the CCDs. The spectral energy distributions of solar
metallicity A0~V, G2~V and K7~V type stars in units of W m$^{-2}$ Hz$^{-1}$,
taken from the BaSeL-2.2 library, are overplotted with (black) dotted lines (the
vertical scale is arbitrary). The dashed (blue and violet) lines correspond to
G2~V and K7~V type stars with $\feh=-1$ (when compared with the dotted
lines they show the effect of a change of \feh). The dashed (red)
line shows the change of the spectrum of an A0 type star due to a change of
luminosity. The solid (orange) line is the N-type carbon star AW Cyg extracted from the
\citet{gunn} library of stellar spectra.}
\label{fig:SED_BBP}
\end{figure}

\begin{figure}
  \begin{center}
    \leavevmode
 \epsfig{file=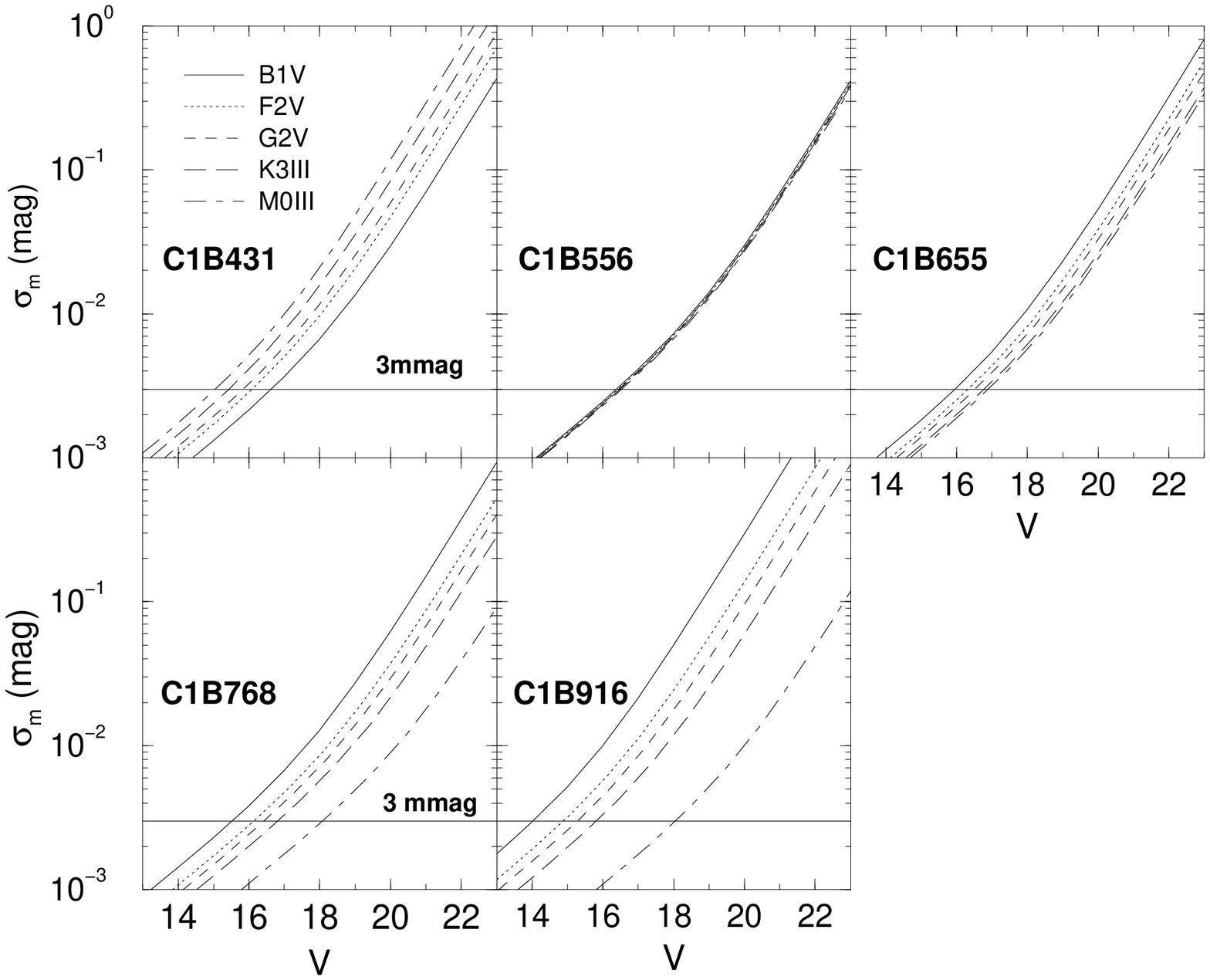,width=1.0\linewidth}
   \end{center}
  \caption{Estimation of the end-of-mission precisions for the five passbands in
  C1B as a function {of~~}$V$, computed according to equation (\ref{eq:eqerror}) and
  taking $\sigma_{\rm cal}=0$~mag. In this paper, the contribution of
  calibration errors to the end-of-mission precision is estimated to be
  3~millimag in every passband (horizontal line). The errors for C1B556 are 
  all nearly the same
  for a given $V$ magnitude because the mean wavelengths of these passbands are
  very similar.}
  \label{fig:errors_BBP}
\end{figure}

The C1B component of the {\Gaia} PS has five broad passbands covering the wavelength
range of the unfiltered light from the blue to the far-red (i.e.\ 400--1000~nm).
The basic response curve of the filters versus wavelength is a symmetric
quasi-trapezoidal shape.  The filters were chosen to satisfy both the
astrophysical needs and the specific requirements for chromaticity calibration
of the astrometric instrument (see Section~\ref{sec:design_PS}). The C1B set of
passbands evolved from the convergence of the \citet{vytasbbp}, \citet{lin2003a}
and \citet{jordibbp} proposals. The specifications of the filters are given in
Table~\ref{tab:filters5}. Figure~\ref{fig:SED_BBP} shows the spectral response
of the passbands.  The estimated end-of-mission precisions, computed as
described in Section~\ref{sec:aperture}, are shown in Fig.~\ref{fig:errors_BBP}.

The Balmer discontinuity and the H$\beta$ line limit the blue and red edges of
the C1B431 filter, respectively. The reddest filter C1B916 is designed to
measure the light between the Paschen jump and the red limit of the sensitivity
of CCDs in the Astro focal plane. The filter C1B655 is centred on the H$\alpha$
line and its width has been optimized together with C1M656 in C1M (see
Section~\ref{sec:section_MBP}). The blue and red limits of the two remaining
filters (C1B556 and C1B768) are consequently set in order to provide full
coverage of the wavelength range in $G$ (i.e.\ avoiding gaps between passbands).
Four passbands are enough for chromaticity calibration, but five passbands are
preferred to four, when the classification and astrophysical parametrization of
stars is considered. Since only four strips of CCDs are available in BBP, the
two broadest passbands (C1B556 and C1B768) are implemented together in one
strip. Hence, the number of observations with these two passbands will be half
as much as with the other C1B passbands.

The current design of the payload foresees no ultraviolet sensitivity for the
Astro/BBP instrument (see Section~\ref{sec:instrument}).  The near ultraviolet
is the most important for stellar classification. The Balmer jump is the feature
in the spectra of B-A-F type stars most sensitive to the temperature and
gravity. It also contains information on the metallicity of F-G-K type stars.
The absence of an ultraviolet passband in BBP is compensated with the inclusion
of a broad UV passband (C1M326) in MBP (see Section~\ref{sec:section_MBP}).  The
classification and parametrization of objects in {\Gaia} is done using the BBP
and MBP measurements together and therefore the lack of a UV passband in BBP
should not be a drawback. However, since the Spectro instrument has a lower
angular resolution than Astro, the combination of BBP and MBP data is not always
possible.  BBP will be the only tool for classification of stars in the crowded
fields with stellar densities larger than $200\,000$--$400\,000$ stars/sq~deg
(see Section~\ref{sec:crowded}). Such stellar densities are found in some areas
of the bulge and of the disk \citep{robin,drimmel2005} and most of these areas 
have low interstellar extinction (such as Baade's window). In
dense areas the trigonometric parallax and the Paschen jump will provide
luminosity parametrization.

\begin{table}
  \caption{{\small Specifications of the filters of C1B implemented
in Astro.}}
\label{tab:filters5}
\begin{center}
\begin{tabular}{lccccc}
\hline
Band       &     C1B431&C1B556&C1B655&C1B768&C1B916 \\
\hline
$\lambda_\mathrm{blue}$ (nm)  & 380  &  492  &  620  &  690  & 866 \\
$\lambda_\mathrm{red}$ (nm)   & 482  &  620  &  690  &  846  & 966 \\
$\lambda_\mathrm{c}$ (nm)     & 431  &  556  &  655  &  768  & 916 \\
$\Delta \lambda$ (nm) & 102  &  128  &   70  &  156  & 100 \\
$\delta \lambda$ (nm) &10,40 & 10,10 & 10,10 & 10,40 &40,10 \\
$\epsilon$ (nm)      & 2,2  &  2,2  &  2,2  &  2,2  & 2,2 \\
$T_\mathrm{max}$ (\%)        &  90  &   90  &   90  &   90  &  90 \\
${n_\mathrm{strips}}$    &   1  &  0.5  &    1  &  0.5  &   1 \\
\hline
\multicolumn{6}{l}{\footnotesize{$\lambda_\mathrm{blue}$, $\lambda_{\rm red}$: wavelengths at half-maximum transmission}}\\
\multicolumn{6}{l}{\footnotesize{$\lambda_\mathrm{c}$: central wavelength$= 0.5(\lambda_{\rm blue} + \lambda_{\rm red})$}}\\
\multicolumn{6}{l}{\footnotesize{$\Delta \lambda$: FWHM}}\\
\multicolumn{6}{l}{\footnotesize{$\delta \lambda$: edge width (blue, red) between 10 and 90\% of $T_{\rm max}$}}\\
\multicolumn{6}{l}{\footnotesize{$\epsilon$: manufacturing tolerance intervals
centred on $\lambda_\mathrm{blue}$ and $\lambda_\mathrm{red}$}}\\
\multicolumn{6}{l}{\footnotesize{$T_\mathrm{max}$: maximum transmission of filter}}\\
\multicolumn{6}{l}{\footnotesize{$n_\mathrm{strips}$: Number of CCD strips carrying the filter: 
C1B556 and C1B768}}\\
\multicolumn{6}{l}{\footnotesize{\hspace{0.4cm}share one CCD strip}}\\
\end{tabular}
\end{center}
\end{table}

\subsubsection {Astrophysical diagnostics}

The response of the C1B431 filter at the shortest wavelength is
asymmetrical, with a red edge that is less steep. This is done to
compensate the shift of the maximum of the response function redward due
to the slope of the quantum efficiency curve and the reflectance curve for
six silver surfaces. As a result, the response function of the C1B431
passband becomes similar to that of the $B$ passband of the {\it UBV}
system. The mean wavelengths of both passbands are also similar:  445~nm
for C1B431 and 442~nm for $B$.

The mean wavelength and half width of the C1B556 response function is very
similar to that of the $V$ passband.  As a result the colour index
C1B431--C1B556 will be easily transformable to Johnson's $B$--$V$ and vice
versa.  Analogously, the C1B768 passband can be easily related with the
Cousins $I$, the SDSS $i'$ and the HST~814 passbands and the colour index
C1B556--C1B768 is transformable to $V$--$I_C$, $r'$--$i'$ and HST~555--814. 
This will facilitate the comparison of the numerous ground-based
investigations in the {\it BV} system and the large number of observations
being done in the far-red passbands with the {\Gaia} results. As can be
seen from Fig.~\ref{fig:SED_BBP}, the differences of the `blue minus
green', `blue minus red' or `blue minus far-red' magnitudes may serve as
a measure of metallic-line blanketing, although with much less sensitivity
than a colour index containing the ultraviolet. In the C1B431--C1B556 vs.  
C1B556--C1B768 diagram the deviations of F-G metal-deficient dwarfs and
G-K metal-deficient giants from the corresponding sequences of solar
metallicity are up to 0.07 and 0.20 mag, respectively.

The combination of the fluxes measured in the C1B655 passband and the narrow
passband C1M656 in MBP form an H$\alpha$ index primarily measuring the
strength of the H$\alpha$ line. The H$\alpha$ index shares the same
properties as the $\beta$ index in the Str\"omgren-Crawford photometric
system. It is an indicator of luminosity for stars earlier than A0 and of
temperature for stars later than A3, almost independent of interstellar
extinction and chemical composition. The same reddening-free index may be
used for the identification of emission-line stars.

The two remaining red and far-red passbands, C1B768 and C1B916, give the
height of the Paschen jump which is a function of temperature and gravity.
Although the maximum height of the Paschen jump is 0.3 mag only, i.e.,
about four times smaller than the Balmer jump, it still provides the
needed information if its height, C1B768--C1B916, is measured with high
accuracy (not lower than 1\%) and this is reachable up to about $V\sim
17-18$ (see Fig.~\ref{fig:errors_BBP}). The colour indices C1B556--C1B768
and C1B556--C1B916 for unreddened late-type stars may be used as
indicators for the temperature. These colour indices (and C1B768--C1B916)
also allow for the separation of cool oxygen-rich (M) and carbon-rich (N)
stars.

The C1M326--C1B431 vs.\ C1B431--C1B556 diagram has the same properties as the
$U-B$ vs.\ $B-V$ diagram of Johnson's system or similar diagrams for other
systems \citep{vytas1992}. Supergiants are well separated from the main-sequence
stars. Metal deficient F-G dwarfs and G-K giants exhibit ultraviolet excesses
up to $0.4$ mag. Blue horizontal branch stars show ultraviolet deficiencies up to
$0.3$ mag, while white dwarfs are situated around the interstellar reddening line
of O-type stars.

\subsubsection{Chromaticity evaluation}
\label{sec:chroma}

Although no refracting optics is used for the astrometric field, the
precise centre of a stellar image is still wavelength-dependent because of
diffraction and its interplay with the optical aberrations of the
instrument. Differential shifts by up to $\sim$10\% of the width of the
diffraction image (i.e., several milliarcsec) may be caused by odd
aberrations such as coma, even though the resolution remains essentially
diffraction-limited. As a result, the measured centres of stellar images
will depend on their spectral energy distributions, and a careful
calibration of the effect, known as \emph{chromaticity}, is mandatory in
order to attain the astrometric accuracy goals. 
The gross spectral energy
distribution of each observed target is therefore needed in the wavelength
range of the astrometric CCDs; moreover, these data are needed with the
same spatial resolution as in the astrometric field. As stated before, the
BBP set of passbands was designed with this requirement in mind,
as well as on astrophysical grounds.

 \begin{figure}
  \begin{center}
    \leavevmode
 \epsfig{file=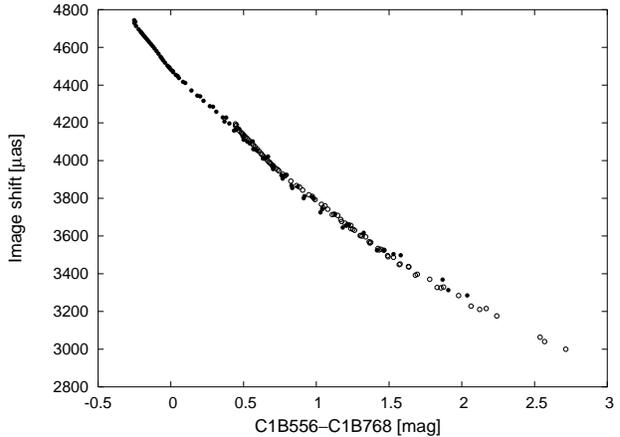,width=1.0\linewidth}
   \end{center}
  \caption{An example of the colour-dependent shifts (chromaticity) 
of stellar images that may be obtained in the astrometric field as 
the result of a coma-like optical aberration (see text for details).
The shift relative to the geometric image centre is plotted versus
a $V-I_c$-like colour index for a range of synthetic stellar spectra
without extinction (filled circles) and for $A_V=2$~mag (open
circles).}
  \label{chrom1}
\end{figure}

 \begin{figure}
  \begin{center}
    \leavevmode
 \epsfig{file=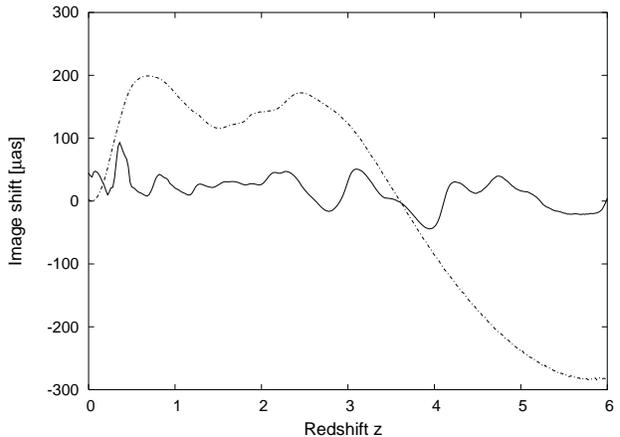,width=1.0\linewidth}
   \end{center}
  \caption{The chromatic shifts of a quasar image before (dash-dotted curve)
and after (solid) correction for the effect calibrated by means of stellar
spectra. A standard quasar spectrum was assumed to be observed at different 
red-shifts, and subject to the same aberrations as in Fig.~\ref{chrom1}.
(The dash-dotted curve has been displaced 4000~microarcsec downwards in the 
diagram to offset the overall shift caused by the coma).}
  \label{chrom2}
\end{figure}

\citet{lin_fil} showed that for the chromaticity calibration,
near-rectangular filters are acceptable and that the choice of the
separation wavelengths is more important than the edge widths. The author
concluded that the use of four broad passbands covering the wavelength
range of the astrometric $G$ passband should be enough to match the
chromaticity constraints (r.m.s.\ contribution to the parallaxes
$<1$~microarcsec).  

Following the design of the passbands in C1B, a more detailed evaluation of
residual chromaticity effects was performed. A worst-case scenario, representing
an extreme amount of coma, was considered where the wavefront aberration consisted of
a third-degree Legendre polynomial in the normalized along-scan pupil coordinate, with
r.m.s.\ wavefront error (WFE) 45~nm. Polychromatic images in $G$ were generated
for a library of synthetic stellar spectra \citep{munari05}, including some
reddened by interstellar extinction.  Image centres were computed through a
modified form of Tukey's biweight formula \citep{nrecipes}, with properties
similar to the maximum-likelihood estimator to be used with the real {\Gaia}
data.

Figure~\ref{chrom1} shows the stellar image shifts for a range of synthetic
stellar spectra plotted versus one of the BBP colour indices.  
For the particular WFE assumed in this example, there is a general shift of 
the image centroid by $\sim\!4$~milliarcsec caused by the coma. However, it 
is only the variation of this shift with the spectral composition that is of 
concern here. The r.m.s.\ variation of the shift versus colour index is 
415~microarcsec.
Using linear regression against the
synthetic stellar C1B counts $\phi_j$ ($j=1\dots 5$, normalized to
$\sum_j\phi_j=1$), the shifts could be reproduced with an r.m.s.\ residual of
7~microarcsec. Further averaging between the $\sim$800 astrometric CCD
observations that are combined in a single parallax will reduce the astrometric
effect of the chromatic residuals by a factor $0.02$--$0.2$ depending on the
degree of correlation among the individual observations, thus leading to a
chromatic contribution to the parallax errors of $0.14$--$1.4$~microarcsec.
Since these numbers are based on a worst-case assumption for the WFE and a
somewhat simplistic calibration model, it is reasonable to expect that a
residual contribution of 1~microarcsec can be achieved, and in particular that
the chosen C1B passbands provide sufficient information on the spectral energy
distribution within the $G$ passband for this purpose. The residual effect is
thus small compared with the statistical errors from photon noise and other
sources even for bright stars.

The chromaticity correction based on an empirical calibration against
stellar BBP fluxes will be less accurate for objects with strongly
deviating SEDs. Prime examples of this are the
quasars, which may exhibit strong emission lines at almost any wavelength
depending on red-shift. Quasars are astrometrically important for
establishing a non-rotating extragalactic reference system for proper
motions; thus chromaticity correction must work also for these objects.
This was investigated by applying the correction derived as described
above from stellar spectra to synthetic quasar images, calculated from a
mean quasar spectrum observed at red-shifts in the range $z=0$ to 6.  
Figure \ref{chrom2} shows the uncorrected image shift as function of $z$
(dash-dotted curve) together with the residual shift after correction (solid
curve). The r.m.s.\ image shift decreases from 170~microarcsec before to
29~microarcsec after correction. The residual curve shows artifacts that
are clearly attributable to the limited sampling of the spectral range.
For example, the negative slopes for $z=2.3$--2.8 and $z=3.1$--3.9
correspond to the red-shifted Lyman-$\alpha$ emission line moving through
the C1B431 and C1B556 passbands, respectively.

With similar assumptions as for the stars concerning the statistical
averaging of the effect when propagating to the astrometric parameters,
the residual effect for the quasars will be a few microarcsec. This is
acceptable since these are mostly faint objects with much larger
photon-statistical errors.

Note that the data in Figs.~\ref{chrom1}--\ref{chrom2} only represent an
example of the image shifts that may occur. The actual behaviour depends
strongly on the shape and size of WFE, which vary considerably across the
field of view, and on the detailed centroiding algorithm.

\subsection{The C1M medium passbands}
\label{sec:section_MBP}

\begin{figure}
\begin{center}
\leavevmode
\epsfig{file=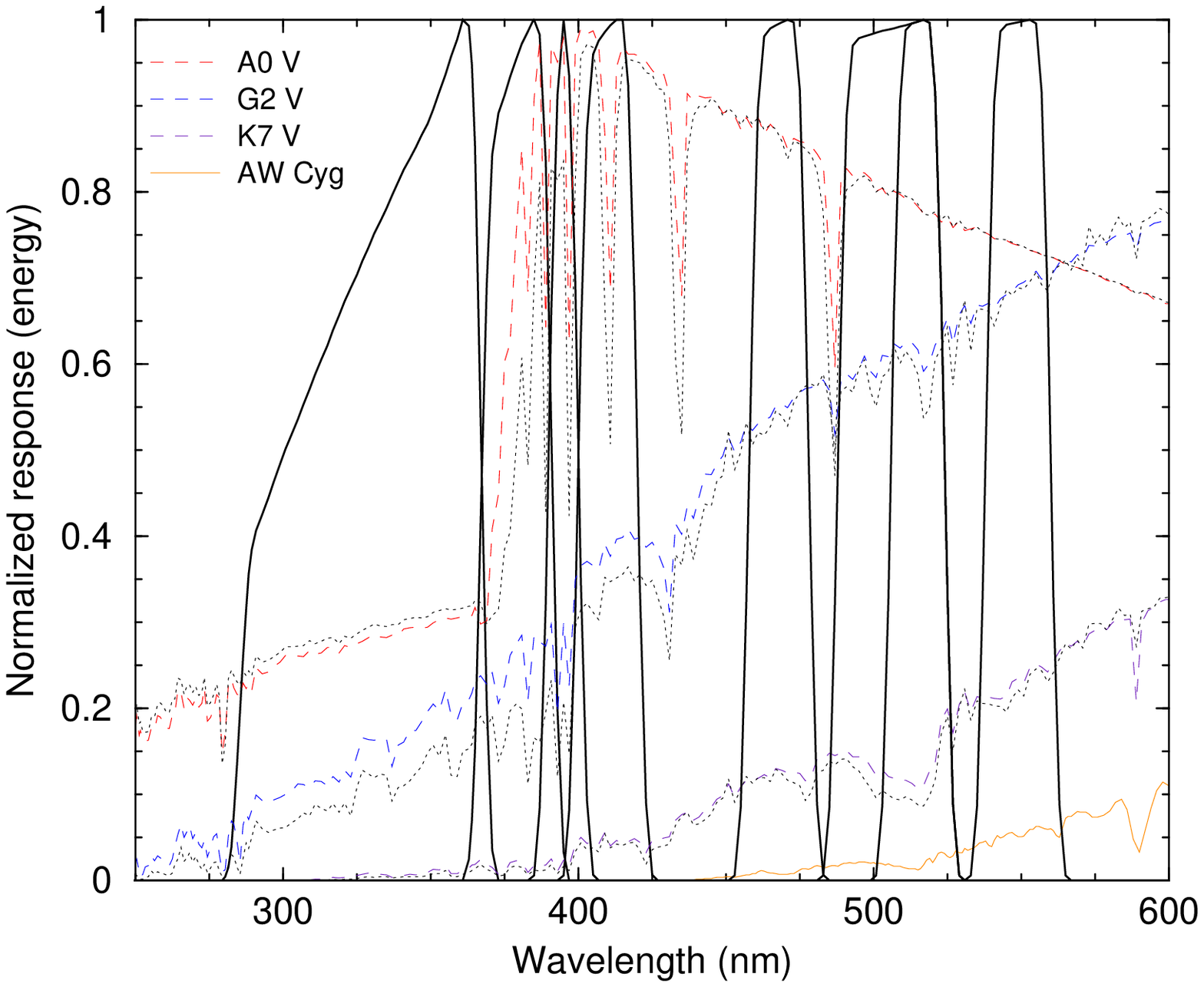,width=1.0\linewidth}
\epsfig{file=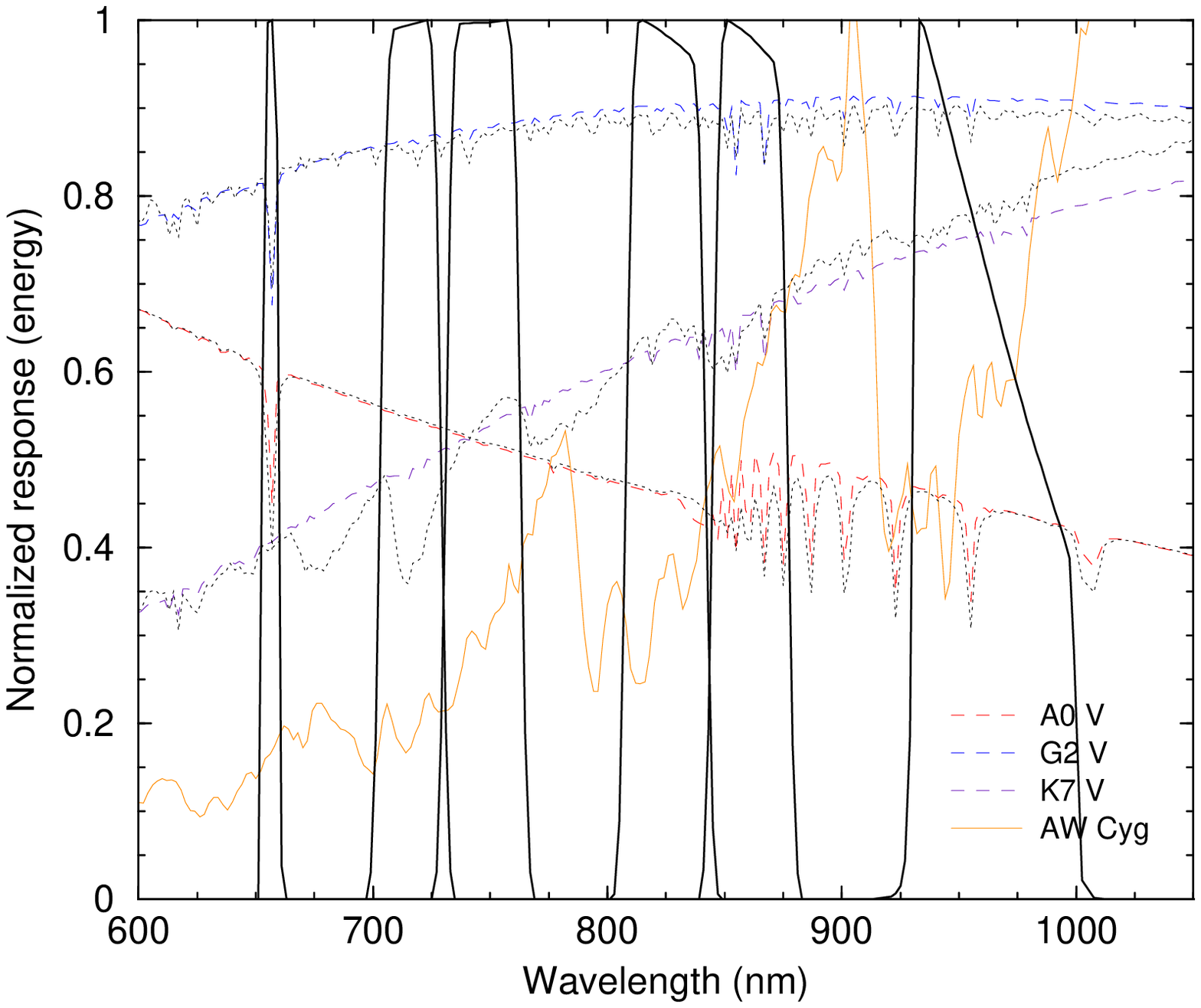,width=1.0\linewidth}
\end{center}
\caption  {Same as Fig.~\ref{fig:SED_BBP} for C1M. Top and bottom
figures show the `blue' and `red' passbands, respectively.}
\label{fig:SED_MBP}
\end{figure}

\begin{table*}
  \caption{{\small Specifications of the filters in C1M implemented
in Spectro.} }
\label{tab:filters6}
\begin{center}
\begin{tabular}{lccccccc}
\hline
Band            &C1M326&C1M379 &C1M395& C1M410& C1M467& C1M506& C1M515 \\
\hline
$\lambda_\mathrm{blue}$ (nm) & 285 &  367  &  390 &   400 &   458 &   488 &   506 \\
$\lambda_\mathrm{red}$ (nm)  & 367 &  391  &  400 &   420 &   478 &   524 &   524 \\
$\lambda_\mathrm{o}$ (nm)    & 326 &  379  &  395 &   410 &   468 &   506 &   515 \\
$\Delta \lambda$ (nm)&  82 &   24  &   10 &    20 &    20 &    36 &    18 \\
$\delta \lambda$ (nm)&   5 &    5  &    5 &     5 &     5 &     5 &     5 \\
$\epsilon$ (nm)     & 2,2 &  2,2  &  2,1 &   1,2 &   2,2 &   2,2 &   2,2 \\
$T_\mathrm{max}$ (\%)        &  90 &   90  &   90 &    90 &    90 &    90 &    90 \\
Type of CCD      &blue & blue  & blue &  blue &  blue &  blue &  blue \\
$n_\mathrm{strips}$ &   2 &    2  &    1 &     1 &     1 &     1 &     1 \\
\hline
Band             &C1M549& C1M656 &C1M716& C1M747& C1M825& C1M861& C1M965 \\
\hline
$\lambda_\mathrm{blue}$ (nm) &   538&   652.8&   703&   731 &   808 &   845 &   930 \\
$\lambda_\mathrm{red}$ (nm)  &   560&   659.8&   729&   763 &   842 &   877 &  1000 \\
$\lambda_\mathrm{c}$ (nm)    &   549&   656.3&   717&   747 &   825 &   861 &   965 \\
$\Delta \lambda$ (nm)&    22&     7  &    26&    32 &    34 &    32 &    70 \\
$\delta \lambda$ (nm)&     5&     2  &     5&     5 &     5 &     5 &     5 \\
$\epsilon$  (nm)   &   2,2&   1,1  &   2,2&   2,2 &   2,2 &   2,2 &   2,2 \\
$T_\mathrm{max}$ (\%)        &    90&    90  &    90&    90 &    90 &    90 &    90 \\
Type of CCD     &  blue&   red  &   red&   red &   red &   red &   red \\
$n_\mathrm{strips}$ &     1&     1  &     1&     1 &     1 &     1 &     1 \\
\hline
\multicolumn{8}{l}{\footnotesize{$\lambda_\mathrm{blue}$, $\lambda_\mathrm{red}$: wavelengths at half-maximum transmission}}\\
\multicolumn{8}{l}{\footnotesize{$\lambda_\mathrm{c}$: central wavelength$= 0.5(\lambda_\mathrm{blue} + \lambda_\mathrm{red})$}}\\
\multicolumn{8}{l}{\footnotesize{$\Delta \lambda$: FWHM}}\\
\multicolumn{8}{l}{\footnotesize{$\delta \lambda$: edge width (blue, red) between 10 and 90\% of $T_\mathrm{max}$}}\\
\multicolumn{8}{l}{\footnotesize{$\epsilon$: manufacturing tolerance intervals
centred on $\lambda_\mathrm{blue}$ and $\lambda_\mathrm{red}$}}\\
\multicolumn{8}{l}{\footnotesize{$T_\mathrm{max}$: maximum transmission of filter}}\\
\multicolumn{8}{l}{\footnotesize{$n_\mathrm{strips}$: Number of CCD strips carrying the filter}}\
\end{tabular}
\end{center}
\end{table*}

 \begin{figure}
  \begin{center}
    \leavevmode
% \centerline{\epsfig{}}
 \epsfig{file=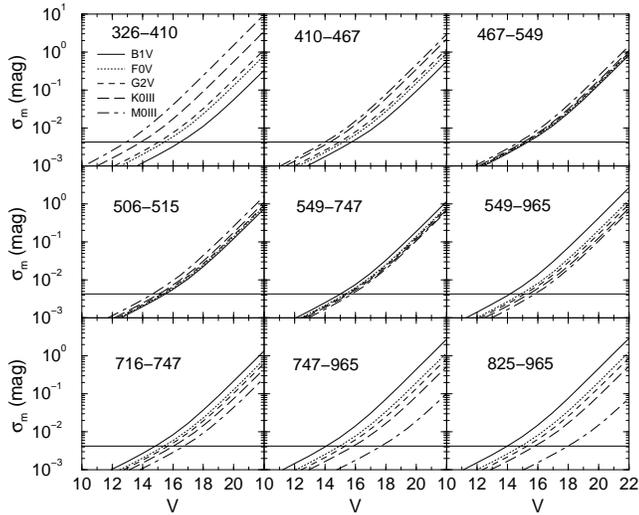,width=1.0\linewidth}
   \end{center}
  \caption{Estimation of the end-of-mission precisions as a function of $V$ for
  several photometric indices constructed from the passbands in C1M, and computed
  according to equation (\ref{eq:eqerror}) with $\sigma_{\rm cal}=0$~mag. In this
  paper, the contribution of calibration error to the end-of-mission precision
  is estimated to be of 3~millimag in every passband.  This is indicated by the
  horizontal line (which here includes a factor of $\sqrt2$).}
  \label{fig:errors_MBP}
\end{figure}

The C1M component of the {\Gaia} PS consists of 14 passbands and evolved from
the convergence of the proposals by \citet{grenon}, \citet{knude},
\citet{vladas1X}, \citet{vytas2004} and \citet{jordimbp}. The guidelines by
\citet{taut2002} for $\alpha$-elements abundance determination were taken into
account.  The basic response curve of the filters versus wavelength is a
symmetric quasi-trapezoidal shape. Their parameters are listed in
Table~\ref{tab:filters6} and the response of the corresponding passbands is
shown in Fig.~\ref{fig:SED_MBP}. Six strips with red-enhanced CCDs are available
for MBP and six red passbands have been designed, implemented as one filter for
each strip, with the only constraint that MBP has to measure the flux entering
the RVS instrument (see Section~\ref{sec:principles_PS}). For the blue passbands
eight filters are implemented on the ten strips with blue-enhanced CCDs. Two
strips have been allocated to each of the two UV passbands to increase the
signal-to-noise ratio (SNR) of the measurements.

The primary purpose of the medium passbands is the classification and
astrophysical parametrization of the observed objects (stars, QSOs, galaxies,
solar-system bodies, etc.). In the case of the stars the goal is to determine
effective temperature \teff, gravity $\log g$ or luminosity $M_V$, chemical
composition [M/H], \afeh\ and C/O abundances, peculiarity type, the presence of
emission, etc., in the presence of varying and unknown interstellar extinction.
Taxonomy classification for solar-system objects and photometric red-shift
determination for QSOs are also aimed for.

\subsubsection {Astrophysical diagnostics}

The filter at the shortest wavelength is C1M326 with wavelengths at half-maximum
transmission of 285 and 367~nm (thus it is of broad passband type although
implemented in Spectro). Below 280~nm, strong absorption lines, metallicity
dependent, are present already in A type stars. The interstellar extinction
increases rapidly below 280~nm and reaches a maximum at 218~nm.
In space the UV passband can be extended down to 280~nm which improves the
determination of {\feh} for F-G-K stars because of the presence of many atomic
lines, ionized or of high excitation. For late G dwarfs, the line blocking in
the extended UV passband is about 2.7 times larger than in the violet
376--430~nm domain. The red side of C1M326 is set at the Balmer jump. The colour
indices C1M326--C1B431 or C1M326--C1M410 give the height of the Balmer jump,
which is a function of {\teff} and {\logg} in B-A-F stars. In F-G-K stars these
colour indices measure metallic line-blocking in the ultraviolet which can be
calibrated in terms of {\feh}.

The C1M379 passband is placed on the wavelength range where the lines
corresponding to the higher energy levels of the Balmer series crowd
together in early-type stars. The integrated absorption in these lines is
very sensitive to {\logg} (or $M_V$). For late-type stars the position of
this passband coincides with the maximum blocking of the spectrum by
metallic lines. Hence the colour index C1M379--C1M467 is a sensitive
indicator of metallicity. Analogs in other photometric systems are the $P$
magnitude in the Vilnius system and $L$ in the Walraven system.

The C1M395 passband is introduced mainly to measure the Ca\,II~H line. The
index C1M395--C1M410 shows a strong correlation with W(CaT*), the
equivalent width of the Calcium triplet measured by the RVS instrument,
corrected for the influence of the Paschen lines. The C1M395--C1M410 vs.\
W(CaT*) may be used as a {\logg} estimator (\citealt{kaltcheva}, Knude \&
Carrasco, priv. comm.).  The C1M395--C1M410 vs.\ W(CaT*) plane is
particularly useful since the effect of reddening on the colour index is
only minor due to the small separation of the two passbands. Additional
uses of the C1M395 passband are in assisting the {\afeh} determination and
in the identification of very metal-poor stars.

The violet C1M410 passband measures the spectrum intensity redward of the
Balmer jump. In combination with C1M326, it gives the height of the jump.
For K--M stars it is the shortest passband which when combined with longer
passbands can provide temperatures and luminosities of solar metallicity
stars in the presence of interstellar reddening (i.e.\ when the stars are
too faint in the ultraviolet).  Its analogs are: $v$ in the
Str\"omgren system, $B_1$ in the Geneva system and $X$ in the Vilnius
system.

The blue C1M467 and green C1M549 passbands measure domains where the absorption
by atomic and molecular lines is minimal.  The flux in these domains correspond
to a pseudo-continuum.
The colour indices C1M467--C1M549, C1M467--C1M747 and C1M549--C1M747 may be used
as indicators of the temperature for stars of all spectral types.  The analogs
of C1M467 are: $b$ in the Str\"omgren system, $B_2$ in the Geneva system, and
$Y$ in the Vilnius system.  The analogs of C1M549 are:  $y$ in the Str\"omgren
system and $V$ in the Vilnius system.

The green C1M515 passband is placed on a broad spectral depression seen in
the spectra of G and K type stars and formed by crowding of numerous
metallic lines. Among them, the strongest features are the Mg\,I triplet
and the MgH band.  The depth of this depression, the intensity of which
reaches a maximum around K7~V, is very sensitive to gravity, being deeper
in dwarfs than in giants. The same passband is also useful for the
identification of Ap stars of the Sr-Cr-Eu type.  The same passband ($Z$)
is used in the Vilnius photometric system.

The C1M506 is much broader and includes the C1M515 passband region. The
combination of both provides an index which is almost reddening free and
its combination with the contiguous pseudo-continuum passbands (C1M467 and
C1M549) provides an index sensitive to Mg abundances and gravity. If the
luminosity is known from parallax, Mg abundances can be determined. The
Ca\,II and Mg\,I spectral features show inverse behaviour when {\feh}
and {\afeh} change \citep{taut2002}, and hence, indices using C1M395 and
C1M515 allow the disentangling of Fe and $\alpha$-process element
abundances.

The narrow passband C1M656 is placed on the H$\alpha$ line.  As mentioned
in Section~\ref{sec:section_BBP}, the H$\alpha$$=$C1B655--C1M656 index is
a measure of the intensity of the H$\alpha$ line, yielding luminosities
for stars earlier than A0 and temperatures for stars later than A3. The
index is most useful for identification of emission-line stars (Be, Oe,
Of, T Tau, Herbig Ae/Be, etc.).

C1M716 coincides with one of the deepest TiO absorption bands with a head at 713
nm \citep{wahlgren}, while C1M747 measures a portion of the spectrum where the
absorption by TiO bands is minimum. So, the index C1M716--C1M747 is a strong
indicator of the presence and intensity of TiO, which depends on temperature and
TiO abundance for late K and M type stars. For earlier type stars, both
passbands provide measurements of the pseudo-continuum. Earlier PS proposals for
MBP considered the inclusion of a filter centred on the TiO absorption band at
781~nm. FoM computations and analysis of the \skpost values showed that the AP
determination improves by more than 10\%, for [Ti/H] and {\teff}, if the
passband is centred on 716~nm instead of on 781~nm.
A similar but narrower passband has been used in the Wing
eight-colour far-red system.

The passband C1M825 is designed to measure either the continuum blueward of the
Paschen jump (hence its limitation at 842~nm for the red side mid-transmission
wavelength) or the strong CN band for R- and N-type stars.  For M stars, C1M825
measures a spectral domain with weak absorption by TiO. The distinction between
M and C stars is realized with all red passbands. At a given temperature, the
fluxes are similar in the C1M747 and C1M861 for O-rich stars (the M sequence)
and for C-rich stars (the C sequence), but very different in the C1M825 and
C1M965 passbands, namely because of strong CN bands developing redward of 787
nm. The separation between M and C stars is possible even if they are heavily
reddened.

Similarly, C1M965 measures the continuum redward of Paschen jump (and in
combination with C1M825 yields the height of the jump) or strong
absorption bands for R- and N-type stars (see Fig.~\ref{fig:SED_MBP}
bottom). Having a passband at these very red wavelengths at the egde of
the CCD QE curve improves the interstellar extinction determination, which
was proven through the FoM computations.

The C1M861 passband, in between C1M825 and C1M965, is constrained by the
wavelength range of the RVS instrument (i.e.\ 848-874~nm) and hence
includes the Ca IR triplet. The measurement of the flux of the star in
this passband will help the RVS data reduction. The index C1M861--C1M965
measures the gravity-sensitive absorption of the high member lines of the
Paschen series.

Finally, the indices C1M825--C1M861 and C1M861-C1M965 are a sensitive
criterion for the separation of M-, R- and N-type stars (see
Fig.~\ref{fig:SED_MBP} bottom).

%%%%%%%%%%%%%%  Photometry precision  %%%%%%%%%%%%%%%%%%%%%%%%%%%%%%%%%

\section{Photometric precision}
\label{sec:phot_precisions}

This section describes the simulation of photometric fluxes that will be
measured by {\Gaia}, the associated magnitude errors and the effect on these
errors of crowded regions and the calibration of the photometric measurements.

\subsection{Simulated photometry}
\label{sec:simulation}

For the simulation of synthetic white light $(G, GS)$, C1B and C1M fluxes and
their corresponding errors, a photometry simulator {\it GaiaPhotSim}\footnote{A
web interface is available at http://gaia.am.ub.es/PWG/} was created. This tool
predicts the number of photo-electrons per unit time (or per CCD crossing) for
every given passband, observed target and {\Gaia} instrument specifications.
{\it GaiaPhotSim} also estimates the associated magnitude and the magnitude
error per transit and at the end of the mission. The inputs are: the instrument
parameters, the SED --- from an empirical or synthetic spectral library --- of
the observed object, its apparent magnitude and radial velocity, the extinction
law and the {\av} value, the brightness of the sky background, and optionally,
the coordinates of the object on the sky.  The coordinates allow the derivation
of the actual number of observations according to the nominal scanning law. The
adopted interstellar extinction law is that of \citet{cardelli} and the ratio of
total to selective absorption can be chosen. The latter has a default value of
$3.1$.  Simulations can be done for single and multiple stars, emission-line
stars, peculiar stars, solar-system objects, QSOs, etc, actually for any given
SED.

To compute the object flux $s_j$, measured in a given photometric passband $j$
and collected after a single CCD crossing, the following ingredients are needed:
the object SED, the assumed interstellar extinction and the extinction law all
resulting in $N(\lambda)$ (in units of photons~m$^{-2}$~s$^{-1}$~nm$^{-1}$), the
transmission profile of the passband $T_j(\lambda)$, the telescope transmittance
$T(\lambda)$, the detector response $Q(\lambda)$ (the CCD quantum efficiency),
the pupil area $A$ and the single-CCD integration time $\tau$. The object flux
is then given by:
\begin{equation} 
\label{eq:eqflux} 
s_{j} [e^{-}] = A
\cdot \tau \cdot \int_{\lambda_{{\rm min},j}}^{\lambda_{{\rm max},j}} \mathrm{d}
\lambda \ N(\lambda)\ T(\lambda)\ \ T_{j}(\lambda)\ Q(\lambda)
\end{equation}

The flux $s_{j}$ is converted into standard magnitudes $m_{j}$ using the
absolute flux calibration of Vega \citep{megessier}, with all its colours set
equal to zero. The SED of Vega has been modelled according to \citet{bessell98}
(a Kurucz SED of $\teff=9550$~K, $\logg=3.95$, $\feh=-0.5$).

By default, the magnitude of the sky background is assumed to be at the
level of the zodiacal light measured in Hubble Space Telescope (HST)
observations and is set to $V=22.5$ mag arcsec$^{-2}$. This assumption is
very conservative for most of the sky. For example, HST measurements show
that the sky background is $V=23.3$ mag arcsec$^{-2}$ at high ecliptic
latitudes.

All simulations in this paper were run with the mean value for the total
number of observations (see Table~\ref{tab:instruments}) and with the
default values for the extinction law and the sky background.

\subsection{Aperture photometry and associated errors}
\label{sec:aperture}

As discussed in Section~\ref{sec:skymappers}, during the observational process
only the pixels in the area immediately surrounding the target source are sent
to the ground in the form of a `window'. In most cases the pixels in the window
are binned in the across-scan direction so that the resulting data consists of 
a one
dimensional set of number counts per sample. Following an `aperture photometry'
approach, it is assumed that the object flux $s_j$ within a given passband $j$
is measured in a rectangular `aperture' of $n_s$ samples within the window. Some
light loss can be produced due to vignetting and/or the finite extent of the
`aperture'. Hence the actual measured flux will be $f_{\rm aper} \cdot s_j$,
where $f_\mathrm{aper} \le 1$.

The end-of-mission magnitude error ($\sigma _{m,j}$) is computed taking
into account: (1) the total detection noise per sample $r$, which includes
the detector read-out noise, (2) the sky background contribution $b_j$
assumed to be derived from $n_b$ background samples, (3) the contribution
of the calibration error per elementary observation $\sigma_{\rm cal}$,
and (4) the mean total number of observations $n_{\rm eff}=n_{\rm obs}
\times n_{\rm strips}$. The resulting magnitude errors are increased by a
20\% ($m=1.2$) `safety margin' to account for sources of error not
considered here (such as a dependence of the calibration error on sky
density and source brightness, see below):
\begin{eqnarray} 
\label{eq:eqerror} 
\lefteqn{\sigma _{m,j} [{\rm mag}] = m
\cdot \frac{1}{\sqrt{n_{\rm eff}}} \cdot \bigg(\sigma^{2}_{\rm cal} +\Big(
2.5 \cdot {\rm log}_{\rm 10}e \cdot } \nonumber\\
 & &  \frac{[f_{\rm aper} \cdot s_{j} + (b_{j} + r^{2}) \cdot n_{s} \cdot
(1+ n_{s}/n_{b})]^{1/2}}{f_{\rm aper} \cdot s_{j}} \Big)^{2} \bigg)^{1/2} 
\end{eqnarray} 

The attainable precisions are shown in Figs.~\ref{fig:errorsG} and
\ref{fig:errors_BBP} where the estimated $\sigma _{m,j}$ are plotted as a
function of $V$ and spectral type for the $G$ and C1B passbands, respectively.
The associated end-of-mission errors for several colour indices formed with the
C1M passbands are shown in Fig.~\ref{fig:errors_MBP}. A precision of $0.01$~mag
is obtained at $V\sim 18$ and $~16$ for the BBP and MBP passbands, respectively.

For high SNR, the end-of-mission magnitude error is limited by $\sigma_{\rm
cal}$.
Following the approach in the {\Gaia} Study Report (\citet{gsr}, p.~264) and
evaluating the number of involved calibration parameters and the number of
available stars for calibration purposes, $\sigma_{\rm cal}$ ranges from 0.3 to
2 millimag depending on the instrument and the passband. Thus potentially
sub-millimagnitude photometric precisions can be achieved at the end of the
mission. However, since there is no detailed calibration model for the
photometric data processing at the moment, we prefer to be very conservative and
we assume an ad hoc calibration error floor of 30~millimag per single
observation, yielding a minimum calibration error of about 3 millimag at the end
of the mission ($\sigma _{m,j} \ge 3$~millimag).

\subsection{Crowded areas, image restoration}
\label{sec:crowded}

An important limiting factor for the photometric accuracy is the distribution of
the background flux around each source. Part of the background consists of
diffuse emission from the general sky background and zodiacal light. In addition
there will be discrete background sources which may be fainter than the {\Gaia}
survey limit ($G_\mathrm{lim}$) and PSF features of neighbouring bright stars
may also contribute to the background. In Astro there is the added complication
that the fields of view from two telescopes overlap on the focal plane and that
the two overlapping fields will be different for different scan directions.

\begin{figure}
  \begin{center}
    \leavevmode
 \epsfig{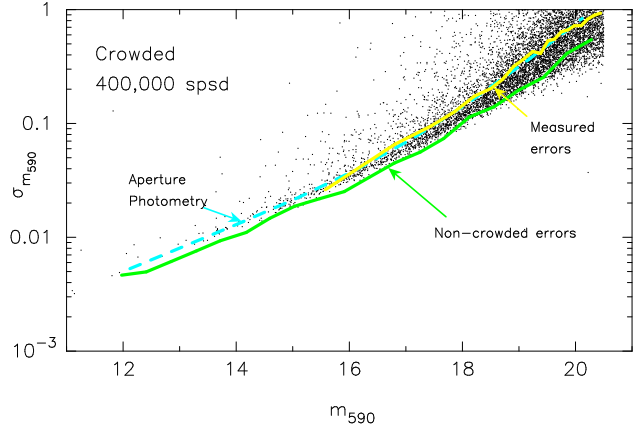}
   \end{center}
	\caption{Various single transit error estimates: the dashed (blue) line is
	for the case of aperture photometry (equation \ref{eq:eqerror}); the solid
	(green) line for the case of PSF fitting photometry using accurate positional
	information from astrometric processing in a non-crowded field, and the
	shorter (yellow) line on the right above the green line shows the same for a
	crowded-field ($400\,000$ stars/sq~deg). The black dots are the formal errors
	from the PSF fitting solution in the crowded case. ($m_{590}$ is the
	magnitude for a representative MBP passband at 590~nm).}
  \label{fig:PSF}
\end{figure}

This background problem becomes particularly severe in the more crowded areas on
the sky (above $\sim10^5$ stars/sq~deg) where the limited resolution of the MBP
instrument will lead to blending of sources in addition to the effects mentioned
above. In this case the data reduction will require a careful deconvolution and
it was shown by \citet{Evans2004} that this is possible for densities up to
about $2$--$4\times10^5$ stars/sq~deg to magnitude 20, by using the accurate
positional information from the astrometric processing and the knowledge of the
PSF in Spectro.
Moreover, the study also showed that the precisions estimated from equation
(\ref{eq:eqerror}) are pessimistic (see Fig.~\ref{fig:PSF}). In non-crowded
regions, PSF photometry yields lower errors than those estimated with aperture
photometry.

The data processing for crowded regions can be improved further if we also
have knowledge of disturbing sources around each object to a magnitude
limit that goes fainter than $G_\mathrm{lim}$. Images providing this
knowledge can in principle be reconstructed from windows that were
obtained for different scan directions. This reconstruction process has
been studied by \citet{Nurmi2005} and \citet{Dollet2005} who present two
different image restoration methods. Both studies show that by combining
the Astro sky mapper windows or a longer AF window one can obtain images
at a resolution of $\sim0.1$--$0.2$ arcsec that go 2--3 magnitudes deeper
than $G_\mathrm{lim}$.  \citet{Nurmi2005} in particular shows that using a
longer AF window one can map the disturbing sources in the immediate
surroundings of objects (within a $\sim2.5$~arcsec diameter) to $V\sim24$
for brightness differences $\Delta V<8$. The image reconstruction will
obviously also be useful in identifying components of multiple stars.

\subsection{Calibration}
\label{sec:calibration}

The photometric precision and accuracy will ultimately be limited by how
accurately the data are calibrated. At an elementary level the goal of the
photometric data reduction is to accurately extract from each CCD image
the object number counts, and to transform these into calibrated fluxes on
a standard scale.

The effects that influence the measured flux can be broadly divided into four
categories: (1) the mirror and filter reflectivity and transmission profiles;
(2) the details of the CCD response, including non-linearities and charge
transfer inefficiency effects; (3) the point spread function and uneven image
motion during the transit; (4) the sky background and disturbance from
neighbouring sources. The calibration process concerns the careful control and
monitoring of these effects, all of which are time dependent.

The {\Gaia} observations are fully self-calibrated and need not rely on
any extant photometric system. The variations in response across the focal
plane, and the variations with time, will be monitored and corrected
taking advantage of the scanning mode in which {\Gaia} will be operated.
For average stellar densities on the sky about 60 and 300 stars per CCD
per second will cross the focal planes of Astro and Spectro, respectively.
This translates to about 660 and 3400 stars per pixel column per 6 hours
(which is the spin period of \Gaia). The same stars are observed
repeatedly in different parts of the focal plane on time scales varying
from hours to weeks to months, up to the mission life time of 5 years.
There will thus be plenty of measurements to perform detailed CCD
calibrations on scales from pixel columns to CCDs.

The photometric system will be defined by the average response for all the
CCDs over the duration of the mission. It is also important to establish a
standard scale, where all non-linearities in the response are corrected.
These effects will be seen as magnitude dependent PSFs, especially for the
brighter sources. A more extensive discussion of the calibration issues
for the photometric data processing can be found in \citet{Brown2005}.

%%%%%%%%%%%%%%  Photometry performances %%%%%%%%%%%%%%%%%%%%%%%%%%%%%%%%%

\section{Photometry performances}
\label{sec:phot_performances}

The role of the photometric system is to astrophysically characterize the
observed objects, and mainly to determine APs for single stars. Thus, prior to
parametrization, a classification is needed. Classification allows the
identification of several kinds of objects (stars, QSOs, etc.). This is done
mostly using the photometry, although parallaxes and proper motions will help
the identification of extragalactic objects. Stellar parametrization with
{\Gaia} is very challenging due to the large range of stellar types (across the
whole HR-diagram) encountered, the complete absence of prior information and,
for the vast majority of stars, the availability of only the 19 BBP and MBP
fluxes. For some 100--200 million stars, parallaxes will be measured at the 10\%
accuracy level.
For the brightest stars, the RVS spectra will also provide information on APs.
Parallaxes and RVS will be used when available, although optimal exploitation of
heterogeneous data requires sophisticated approaches. Other challenges are
presented by: the presence of degeneracies (where, e.g., a set of fluxes may
correspond to more than one type of star, see also Section~\ref{sec:degen}); the
very large cosmic variance compared to limited data; the inevitable presence of
strange objects; the problem of accurately determining `weak' APs (\feh,
\logg) in the presence of dominant variance from `strong' APs (\teff, \av).

This section presents two different evaluations of the performance of the
{\Gaia} PS with respect to the determination of APs for single stars. One
approach is through the analysis of the posterior errors introduced in
Section~\ref{sec:fom}, taking into account photometry and parallax information.
The other approach is through the analysis of results obtained with
parametrization algorithms specifically designed for the AP determination. 
In both cases we assume that the differences among the measured fluxes of the
stars, and hence the changes in their spectra, are only due to differences of
the APs $p_k$ and the errors. We also assume that the SEDs, and their change as
a function of the parameters $p_k$, match the true stars. In other words, cosmic
dispersion (the effects of rotational velocity, magnetic fields,
element-to-element chemical composition differences, differences in the
interstellar extinction law, unresolved binarity, and so on) and inaccuracies in
the SEDs libraries (absence of chromospheres and some opacity sources, NLTE
effects, simplified geometry, etc) are not taken into account. In addition, for
both approaches we assume that the noise model described in
Section~\ref{sec:aperture} is correct. Therefore, both the posterior errors as
well as the errors estimated from the parametrization algorithms are, in
principle, optimistic. The posterior errors, which evaluate locally the
sensitivity of the PS with respect to APs, are more optimistic than the results
from the parametrization algorithms because global degeneracies are assumed not
to exist (see Section~\ref{sec:degen}). On the other hand, the errors on the APs
estimated from the parametrization algorithms reflect a combination of the
capabilities of the PS and of the algorithms themselves. Thus a large estimated
error may mean a poor PS performance or a poor efficiency of the algorithm, or
both. The parametrization algorithm discussed in Section~\ref{sec:nn} only
considers the C1M passbands; information from BBP and parallax is not included.
Hence, the estimated errors tend to be pessimistic.

By the end of the {\Gaia} mission improved stellar atmosphere models, improved opacity
sources, improved treatment of convective atmospheres, dusty atmospheres,
geometry, NLTE effects, etc. will be available. This, supplemented with real
data where appropriate, will allow us to minimize one source of inaccuracies.

\subsection{Posterior error, \skpost, estimations}
\label{sec:sigmapost}

\begin{figure}
\begin{center}
  \leavevmode
\epsfig{file=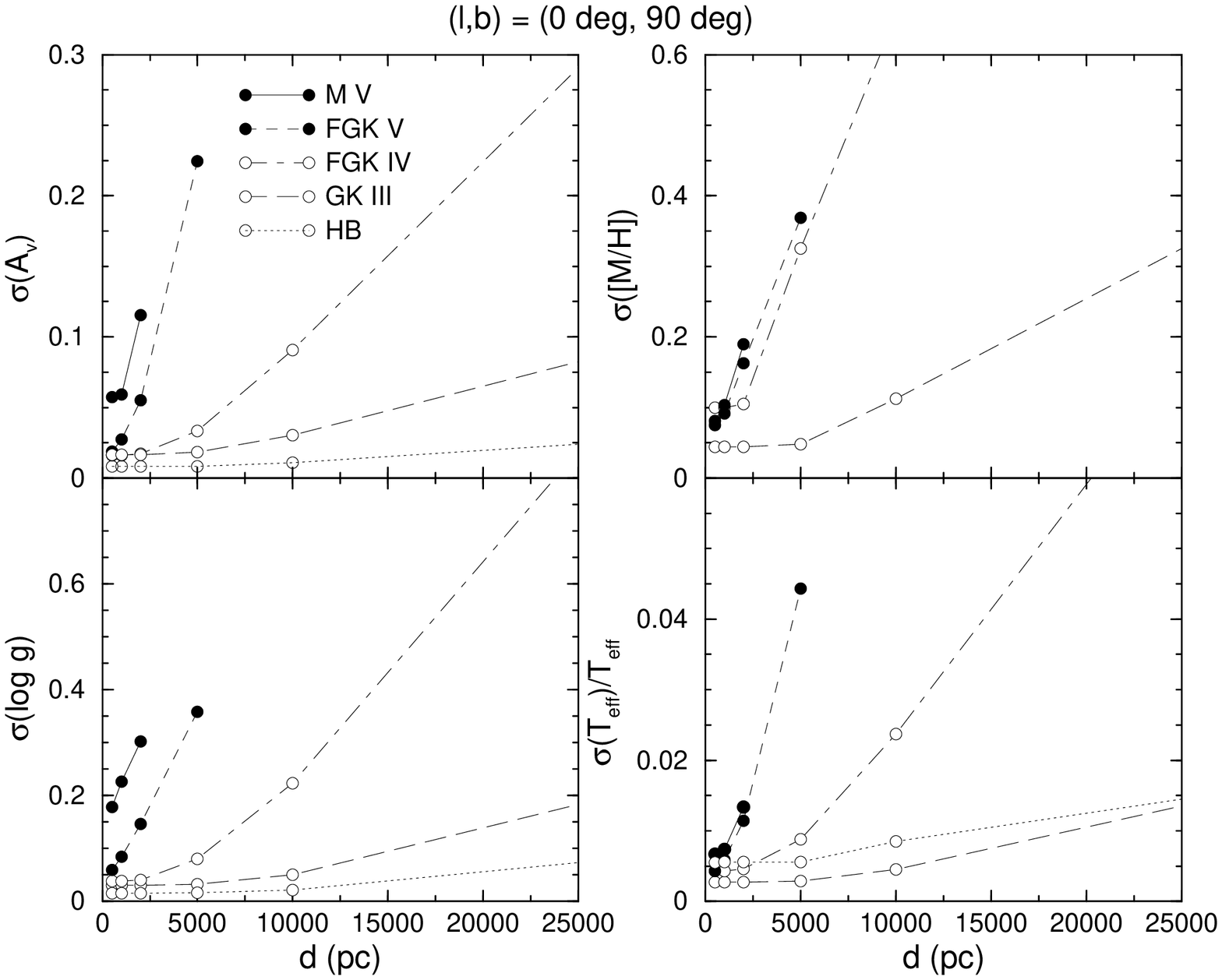,width=1.0\linewidth}
\epsfig{file=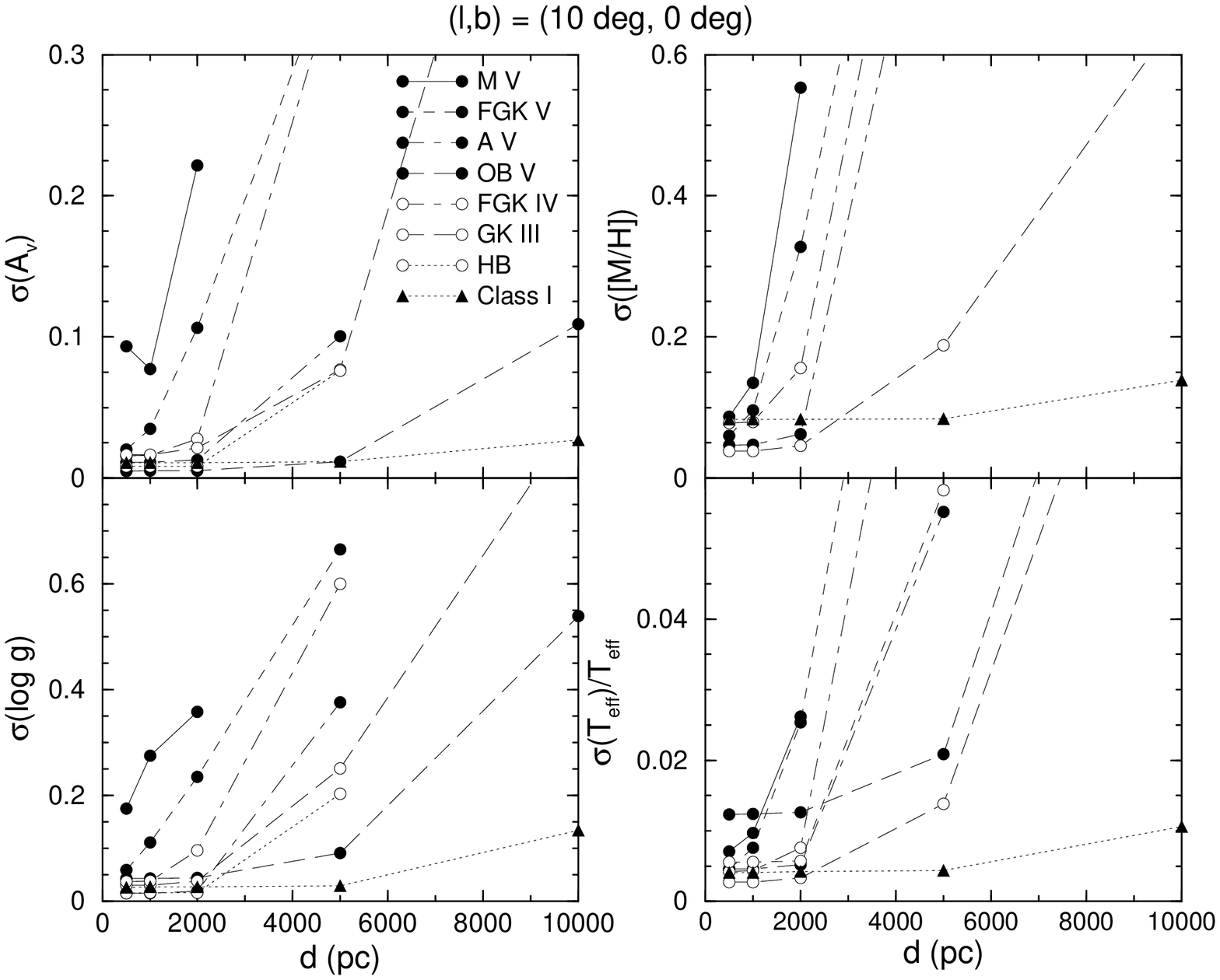,width=1.0\linewidth}
\epsfig{file=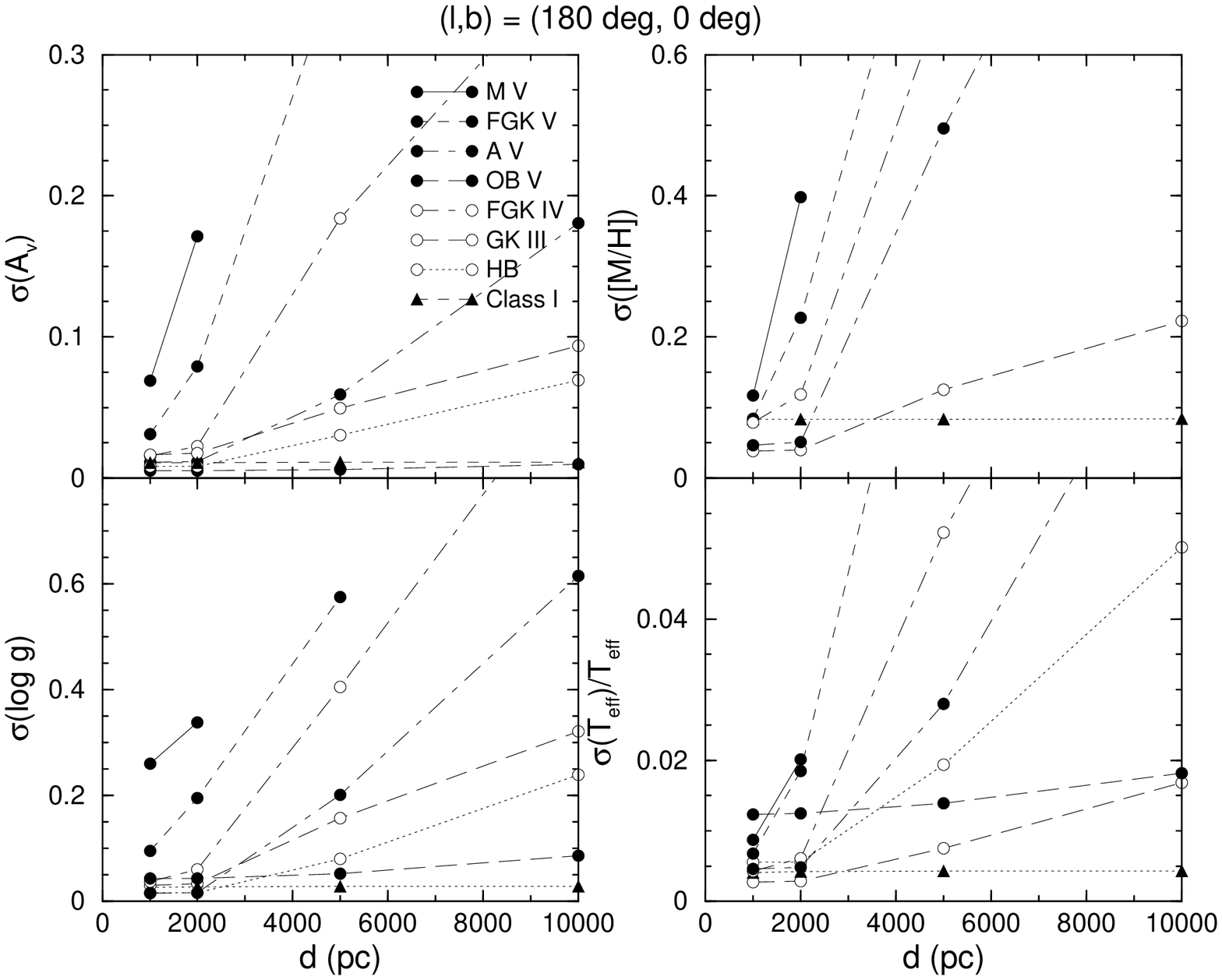,width=1.0\linewidth}
\end{center}
\caption{Predicted (posterior) errors for the main astrophysical parameters 
in three Galactic directions based on G, C1B, C1M and parallax data. The 
following assumption holds for \av: in the Galactic pole direction it 
reaches $0.3$~mag at
1~kpc and is constant from this distance on; toward the Galactic centre it
ranges from $0.3$~mag at 500 pc to 10~mag at 10~kpc; toward the anti-centre it
varies from $0.3$~mag at 500 pc to $3.5$~mag at 5~kpc and is constant for large
distances. The figure legends indicate the meaning of the lines, where
`Class~I' means luminosity class~I and `HB' means horizontal branch stars.
    }
  \label{fig:sigma_post1}
\end{figure}

The posterior errors on the APs, {\skpost}, for the baseline C1B and C1M PS have
been computed as explained in Appendix~\ref{ap:fom} and are presented here. The
targets from Section~\ref{sec:test_pop} have been grouped by spectral type,
luminosity class and Galactic direction and their mean
$\sigma_{\teff,\mathrm{post}}/\teff$, $\sigma_{\av,\mathrm{post}}$,
$\sigma_{\logg,\mathrm{post}}$ and $\sigma_{\feh,\mathrm{post}}$ are shown in
Fig.~\ref{fig:sigma_post1} as a function of distance from the Sun (i.e.\ as a
function of apparent magnitude). The figure shows the error predictions based on
the BaSeL2.2 SED library and the combination of the white light, BBP and MBP
fluxes with parallax information. The fluxes measured in the different passbands
are assumed to vary only due to changes in \teff, \av, {\logg} and \feh. Using
other SED libraries leads to slightly, but not significantly, different \skpost\
values. The largest differences occur for cool stars ($\lesssim$4000~K) for
which the differences among synthetic spectral libraries are large. 
For these stars, discrepancies among theory and observations also exist. See
\citet{arunas} for a comparison of observed colour indices and predictions by
the PHOENIX, MARCS and ATLAS model atmosphere codes for late-type giants.

For a given group of spectral types and luminosity classes, the posterior errors
increase with the distance to the Sun, because the stars become fainter and the
measurement errors of their fluxes increase. In the Galactic plane the stars are
also increasingly reddened (and fainter) with distance which also leads to
larger errors. Given a distance and a group, the posterior errors increase from
the Galactic pole direction to the Galactic centre direction due to the increase
of interstellar extinction. For a given distance and direction, the errors are
larger for dwarfs than for giants and supergiants due to the different absolute
magnitudes.

The lower limit on the {\skpost} values reflects the minimum photometric
error due to the contribution of calibration errors. As discussed in
Section~\ref{sec:calibration} we have conservatively assumed an ad hoc
error floor of $\sim$3~millimag at the end of the mission.

Bulge stars are generally faint because of their distance and the high
extinction toward them. However, there are areas of low extinction, such
as Baade's window, but there the stellar density is very high and
MBP measurements will not be available. In those cases the derived
stellar parameters, based only on $G$, BBP photometry and parallaxes, are
less precise. Nevertheless, the estimated $\sigma_{\av,\mathrm{post}}$
ranges from $0.2$ to $0.6$~mag, $\sigma_{\teff,\mathrm{post}}/\teff$ from
1 to 10\% and $\sigma_{\logg,\mathrm{post}}$ from $0.1$ to $0.5$~dex,
depending on the spectral type. In the high extinction areas, the lower
stellar densities allow for MBP measurements and even though the stars are
very faint, the estimated precision of the APs is
still acceptable: $\sigma_{\av,\mathrm{post}}$ ranges from $0.4$ to
$1.0$~mag, $\sigma_{\teff,\mathrm{post}}/\teff$ from 2 to 15\% and
$\sigma_{\logg,\mathrm{post}}$ from $0.3$ to $0.7$~dex.

\begin{figure}
\begin{center}
    \leavevmode
 \epsfig{file=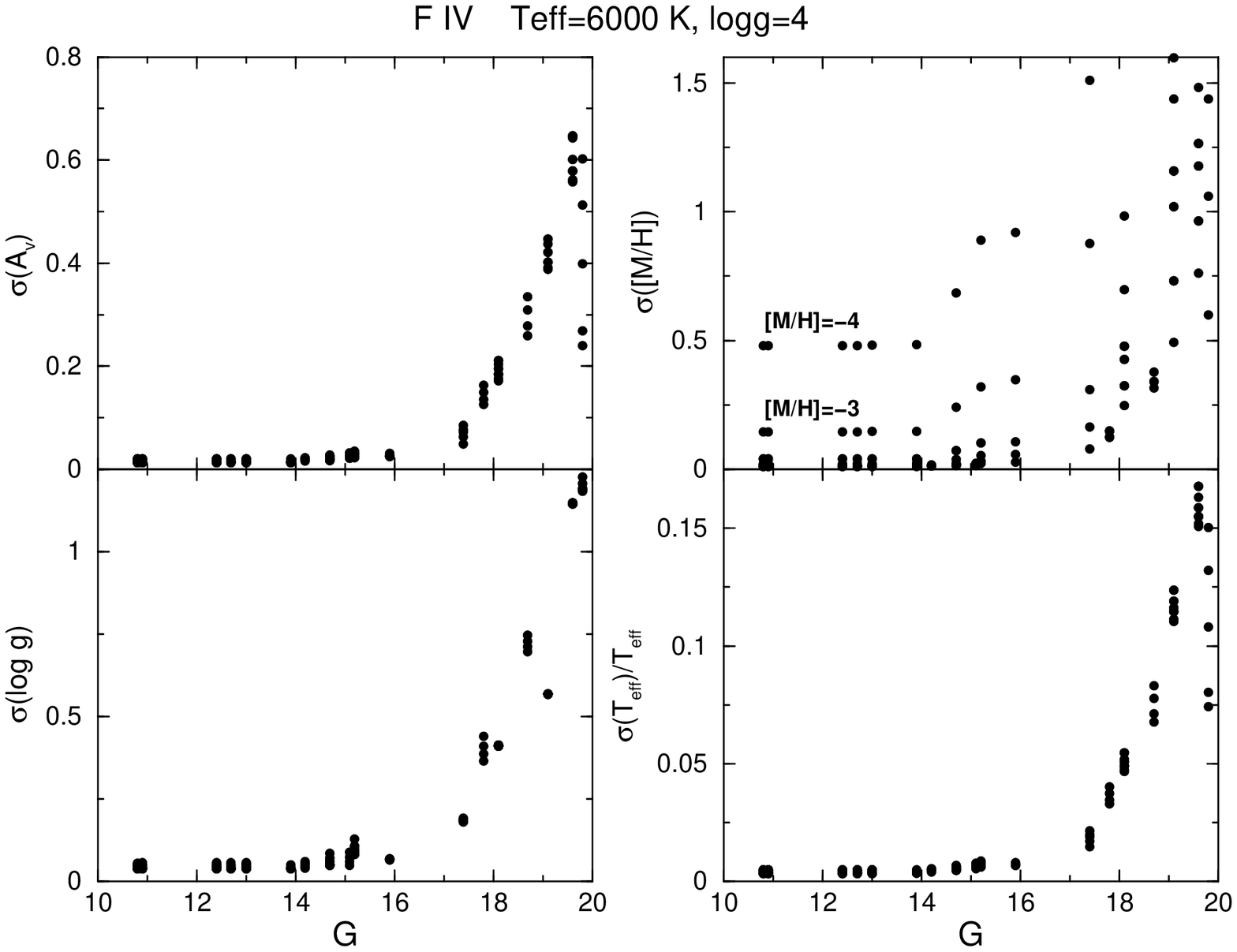,width=1.0\linewidth}
 \epsfig{file=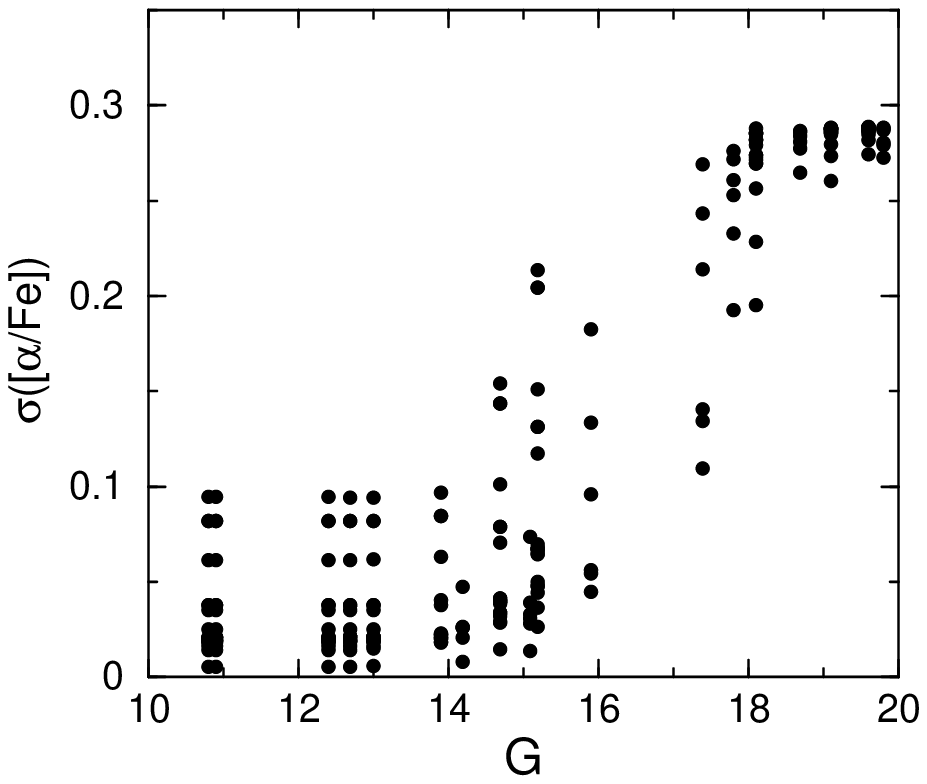,width=0.49\linewidth}
 \epsfig{file=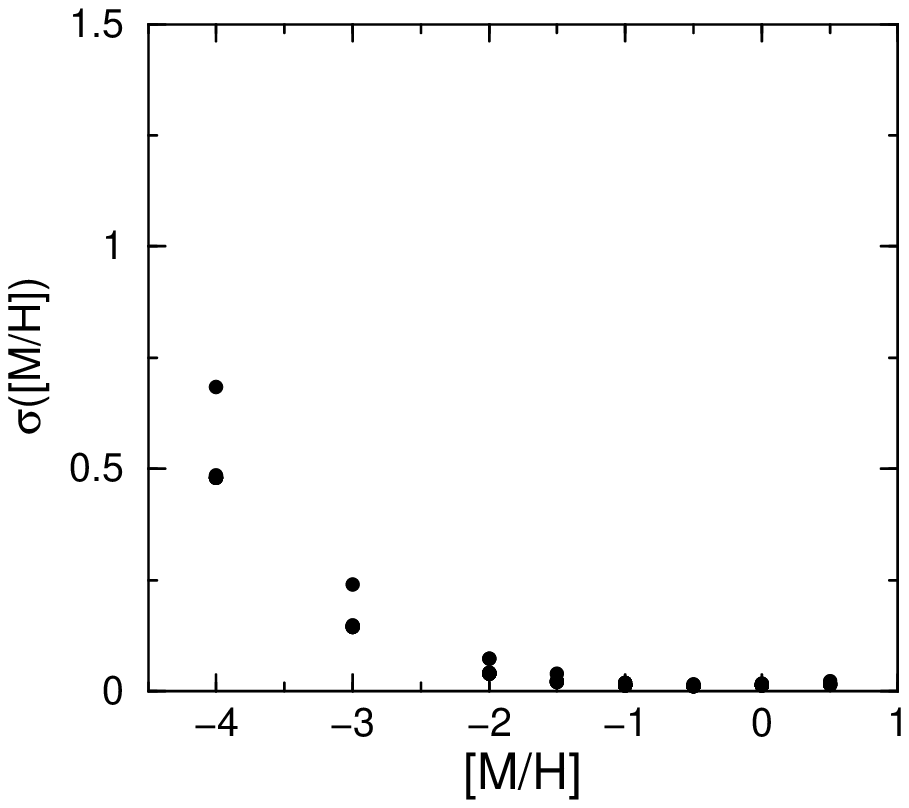,width=0.49\linewidth}
\end{center}
  \caption{{\small 
Predicted (posterior) errors for the astrophysical parameters as a function of
apparent $G$ magnitude and [M/H] for F-type subgiants. The top four panels show
the results based on the BaSeL2.2 SED library. 
In the top right panel the metallicity increases from $\feh=-4$ at the top to
$\feh=+0.5$ at the bottom for each vertical set of dots. The bottom left panel
shows the predicted errors for {\afeh} based on the NextGen2 SED library. The
bottom right panel shows the error on {\feh} as a function of {\feh} for stars
brighter than $G=15$ (based on BaSeL2.2 SEDs).
}}
\label{fig:sigma1}
\end{figure}

Figure~\ref{fig:sigma1} shows the {\skpost} values as a function of apparent $G$
magnitude and {\feh} for F-type stars representative of 
isochrone turn-offs for the old-disk, thick disk and halo stars. This example
has been chosen because F-type stars at the turn-off are key targets for many
Galactic topics (see Table~\ref{tab:tracers}).  The performance of the PS is
excellent up to about $G=18$, even for the most metal-poor stars, and decreases
for faint magnitudes. The determination of {\afeh} is only possible up to
$G\approx16$, while at $G\approx18$, the measurement errors are large enough to
make the changes in SEDs indistinguishable which means that $\sigma_{\rm
\afeh,\mathrm{post}}\sim \sigma_{\afeh,\mathrm{prior}}=0.3$~dex.

As discussed in the introduction of this Section, the \skpost values are
optimistic by construction and from the figures above they may seem too
optimistic compared to the achievable precisions with the currently available
PSs. However, they are not necessarily overly optimistic. {\Gaia} photometry
will provide measurements in 5 broad and 14 medium passbands and the $G$ white
light passband, to which the information contained in the parallaxes can be
added. There is no precedent for deriving stellar astrophysical parameters in
this way.

The {\skpost} here represent the maximum precision that can be obtained.  We
estimate that the main uncertainty on the actual values comes from the imperfect
knowledge of the SEDs of the true stars and not from the design of the
photometric system itself.

\subsection{Estimating the astrophysical parameters}
\label{sec:nn}

The predicted errors discussed above are evaluated locally and hence they are
based on 
the assumption that global degeneracies (e.g., the confusion between a reddened
hot star and an unreddened cool dwarf) are already accounted for before the APs
are estimated. We now discuss the performance of the {\Gaia} PS in a more
realistic setting where the 
APs of each source have to be derived from the measured photometry in the
presence of global degeneracies.

\subsubsection{Methods for astrophysical parametrization}

Numerous statistical methods are available for estimating parameters from
multidimensional data. The most appropriate are supervised pattern recognition
methods, which involve a comparison of the observed `spectrum' (the set of
observed filter fluxes) with a set of template spectra with known APs (the
`training grid').

Minimum distance methods (MDMs) make an explicit comparison with every spectrum
in the training grid to find the `closest' match according to some distance
metric  \citep[e.g.][]{katz98}. Because only discrete APs of templates in the
training grid can be assigned to the observed spectrum, the precision is limited
by the `AP resolution' in the grid. This may be partially alleviated by
averaging over the APs of the nearest neighbours or by interpolating
between them (e.g.\ \citealt{malyuto05,bridzius}).

Instead of making an explicit comparison with every template spectrum, we can
model the mapping between the observed data space, ${\cal D}$, and the AP space,
${\cal S}$, with a regression function $\vect{p} = \vect{f}({\vecphi};
\vect{w})$, i.e.\ the APs \vect{p} are a function of the fluxes {\vecphi} and a
set of free parameters \vect{w}.  We then solve for the free parameters \vect{w}
via a numerical minimization of the errors in the predicted APs of the training
data.  As both ${\cal D}$ and ${\cal S}$ are multidimensional, and because
${\cal D} \rightarrow\ {\cal S}$ is an inverse mapping (and hence may show
degeneracies), the problem is complex \citep{bailer-jones03,bailer-jones05a}.
Nonetheless, it offers advantages over the direct template matching, such as
less dependence on a dense template grid, provides continuous AP estimates and
much faster application times to new data (because the mapping function is
learned in advance).  There are many ways to implement a multidimensional
regression, including neural networks (NNs) and adaptive splines (see e.g.\
\citealt{hastie01}).

Various methods have been tested by the {\Gaia} classification and photometry
working groups and each have their advantages and disadvantages.
For more details see \citet{brown03}.  In the rest of this Section we focus on
one particular method. It must be stressed that this is just a preliminary
investigation into AP estimation with {\Gaia} and is restricted to just the C1M
passbands. 

\begin{figure}
\centering
\includegraphics[scale=0.5]{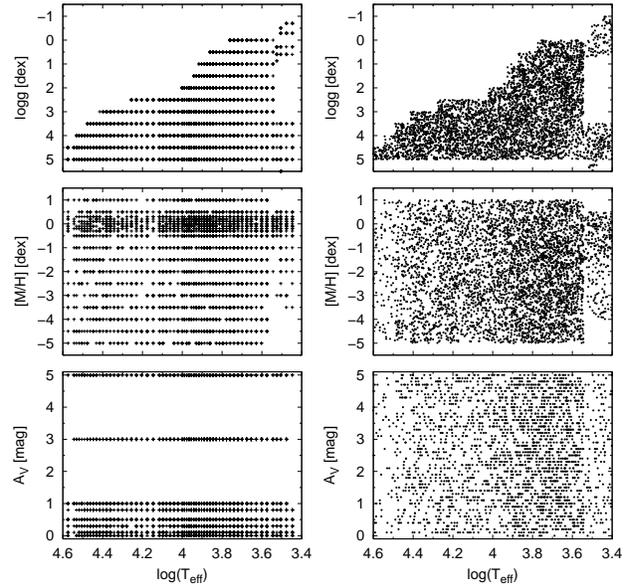}
\caption{The distribution of the astrophysical
parameters in the training (right) and test set (left). In total there
are $61\,941$ stars (AP combinations) in the training set and 9229
different stars (each represented by ten different noisy filter fluxes
for a given magnitude) in the test set.
\label{parset}}
\end{figure}

\subsubsection{A regression model}

We applied a feed-forward NN to a set of simulated photometry in the medium-band
C1M component of the {\Gaia} PS. The training grid consists of $61\,941$ simulated
`spectra' showing variance in the four APs {\teff} (2000~K to $50\,000$~K),
{\feh} ($-5.0$ to $+1.0$), {\logg} ($-1.0$ to $+5.5$), \av, the line-of-sight
interstellar extinction (0.0 to $+5.0$).  The source spectra were taken from the
BaSeL2.2 library. The distribution over the APs is discretized and non-uniform
(Fig.~\ref{parset}, right panel).  We used the {\sc statnet} NN code
\citep{bailer-jones98a,bailer-jones2000} with two hidden layers each comprising 
25 hidden nodes.
{\sc statnet} is trained with a conjugate gradient optimizer with weight decay
regularization to avoid over-fitting. We assess the performance by applying the
network to a separate set of data; the distribution of the APs in this `test
set' is shown in Fig.~\ref{parset} (left panel). In total there are 9229
different stars (AP combinations) in the test set, each represented by ten
different noisy passband fluxes for a given magnitude. Separate models were
trained and tested at magnitudes $G$=15, 18 and 20, i.e.\ with the appropriate
amount of simulated noise, from which we can assess performance as a function of
magnitude.
All combinations of individual values of the four APs and the apparent $G$
magnitude have been used without regard to the reality of such combinations. 
Moreover, we do not build in any prior information
concerning the probability of occurrence of the various AP combinations.

\subsubsection{Results}

\begin{figure}
\centering
\includegraphics[scale=0.6]{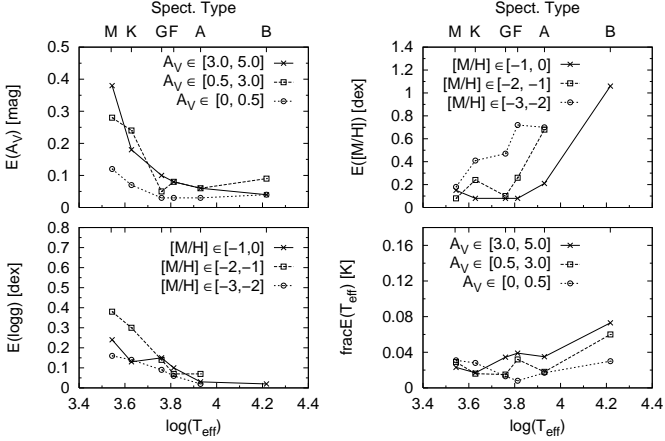}
\caption{Parametrization errors (as defined in equation \ref{eq:err}) for the four
astrophysical parameters (APs) as a function of \teff\ for different subsets of
\av\ and \feh.  For a given \teff\, a data point lies at the average of a
representative temperature interval.  Other than for the \logg\ error plot,
results are shown averaged over the range \logg\ $\in$ [3.5, 5.5] dex, although
the results are not significantly different for giants.  For \av\ and \teff\ the
results are shown for \feh$\in$ [$-$1, 0] dex, for metallicity and gravity, \av\
is limited to [0, 0.5] mag.  The errors for \teff\ are the fractional errors.
While we show results for these limited AP ranges for clarity, it is important
to realize that the model was trained and tested on the full range of APs shown
in Fig.~\ref{parset}: there was no prior restriction on the APs of a star
presented to the model, even though several combinations of $G$ and APs may not
exist in the Galaxy. All results for the C1M passbands for stars at $G$=15 mag
using simulated photon fluxes are representative of the end-of-mission.
\label{g15res}}
\end{figure}

The results are summarized in terms of the average absolute errors
\begin{equation} 
  \label{eq:err}
  E = \frac{1}{N} \cdot \sum_{p=1}^N \left|{\rm computed}(p) - {\rm true}(p)\right|\,,
\end{equation}
where $p$ denotes the $p^{\rm th}$ spectrum (star) and computed($p$) is
the parametrization output provided by the network.  The errors for \teff\
are given as the fractional errors.  It is well known that the error in
any AP for a given star depends significantly not only on the value of
that AP, but also on the value of the other APs.
Therefore, reporting a single error for each AP by averaging over the full
range of the other APs has little meaning and also depends strongly on the
AP distribution in the test set.  We therefore present errors averaged
over selected narrow ranges of APs.  This is only for reporting purposes.
When the network is presented with a star to classify, it has no prior
knowledge of its APs (not even that it is restricted to the range in the
training set).

Representative plots of the AP errors are shown in Fig.\,\ref{g15res} 
for $G$=15.
This plot is
for dwarf stars, i.e.\ {\logg} $\in$ [3.5, 5.5] dex, although the results
are not significantly different for giants.  A complete overview of the
results for all stellar types is given in \cite{willemsen05a}.

For hotter stars (A, B type stars) the errors in {\feh} and {\teff} are
larger than those for cooler stars (G and K type) while the {\av} and
{\logg} errors are smaller. This behaviour is expected and reflects
what we know of how the APs are expressed in the SED as a function of the APs
(in particular \teff). 
For stars of G type and earlier, the extinction is determined quite  
well. For example for A stars we find errors of 0.03 to 0.06 mag for   
$G$=15 and 0.07 to 0.15 mag for $G$=18 mag. For cooler stars (K and M),
the performance degrades slightly, especially for highly extinct cool
stars. 

Metallicity determinations are possible for cool and intermediate
temperature objects (M to F type) with precisions ranging from $0.1$ to
$0.4$ dex, even at low metallicities. For the same stars,
{\feh} can be determined to $0.1$--$0.2$ dex down to $-2$ dex
metallicity. Here, only stars with {\teff}$\ge$3500~K have been
included. For stars below $\sim$4000~K, the estimated uncertainties
may be unrealistic due to the inaccuracies of the current atmosphere models,
which in this work are assumed to match the reality, as discussed
previously.

The precision of the gravity determination depends on the temperature, but
is basically independent of metallicity. We find errors of E(\logg) =
$0.08$ to $0.4$ dex for these stars at $G=15$ mag and $0.2$ to $\sim 1.0$
dex at $G=18$ mag. For hot stars (A, B type), we find errors of $\sim$0.1
dex or better, even for $G=18$ mag. This will provide good discrimination
between distant halo HB stars and nearer A dwarfs in those cases where the
parallaxes are poor.

Temperatures can be determined to between 1\% and 5\% for M to A type
stars ($G=15$ mag) and to precisions of 2\% to 11\% at $G=18$ mag.  There
is some dependence on the extinction. For lower values of the extinction
(\av $\in$ [0, 0.5] mag)  
we find temperature errors of
$\mathrm{E}(\teff)\leq4$\% for ($G=15$) and $\leq5$\% ($G=18$) for all
types of stars.

We performed additional tests to determine the {\afeh} abundance using the
NextGen2 SED library. This increases the dimensionality of the AP space
(from four to five) and thus its complexity. We used the same NN
approach and the C1M passbands. We found errors of
$\mathrm{E}(\afeh)\sim0.1$ dex for low to intermediate temperature stars
at $G=15$ mag. Our results also suggest that precisions of $0.2$ dex
should be possible for $G\lesssim 16.5$ mag.  See \citet{willemsen05b} for
more details.

Comparative tests were also made with MDM and Radial Basis Function Neural
Networks (RBFNNs) using the same data. In general, feed-forward NNs
yielded the best results, followed by MDM and then RBFNN.

\subsection{Discussion}

These preliminary results, based on the C1M passbands, demonstrate that an
automated `bulk' determination of astrophysical parameters is possible.

The posterior errors
predicted from the sensitivity of the SEDs to changes in APs are lower
than the errors estimated from the NN results because the
former are immune to global degeneracies (by design), whereas the NN
performance is degraded by its failure to explicitly deal with
degeneracies (which produce a non-uniqueness in the mapping it is trying
to solve). The NN performance may be degraded by inefficiencies in the 
formulation of the problem via multidimensional regression. 
Yet, the posterior errors are based on C1B, C1M and parallax data, while
the errors with the NN are estimated using only C1M passbands. Taking all
into account, Figs.~\ref{fig:sigma_post1}, \ref{fig:sigma1} and \ref{g15res}
show an overall agreement.

The actual set of training data we will use for the {\Gaia} data
processing will be based on improved stellar atmosphere models 
supplemented with real data.
However, even then the training data are
unlikely to represent the full range of cosmic variance in the APs which
{\Gaia} will encounter.

The performance predictions from the present implementation of the NN
discussed above are pessimistic in several ways. First, only the
C1M passbands were used. Including data from the five C1B
passbands will improve the results (for instance, the index C1B655-C1M656
helps to break degeneracies among \teff, {\logg} and \av). Second, the use
of the parallax, when available, will also help as it provides information
on the intrinsic luminosity (for known apparent magnitude and estimated
extinction) which will help solving degeneracies between {\logg} and
chemical composition. Initial tests of including the parallax as an
additional network input confirm that this improves the estimation of
the gravity, especially for those stellar types for which {\logg} is 
otherwise poorly constrained (i.e.\ for intermediate and low-temperature 
stars). It also improves the metallicity estimates \citep{willemsen05c}.
This decrease of global degeneracies makes the estimated errors from NNs
more similar to the predicted posterior errors. Third, for the brightest
stars, the RVS provides high SNR spectra for all types of stars. This
should improve the AP estimates. Finally, and most importantly, we know
that the method we have used is limited in a number of ways. For example,
the data to AP mapping must be inferred based only on the training data.
Explicitly providing information on the sensitivity of passbands or passband
combinations to APs will help. Likewise, the NNs do not yet
deal with AP degeneracy. This is a particular problem at low SNR. Overall,
the predicted PS performances from Section~\ref{sec:nn}
are rather conservative estimates of what will ultimately be possible with
\Gaia.

%%%%%%%%%%%%%%   Implications in science  %%%%%%%%%%%%%%%%%%%%%%%%%%%%%%%%%

\section{Science implications}
\label{sec:galaxy}

As emphasized a number of times, the main scientific goal of {\Gaia} is the
quantitative description of the chemical and dynamical evolution of the Galaxy
over its entire volume. This is only possible if physical properties of stars,
through chemical abundances and ages, are analyzed together with kinematics and
distances. To determine ages and abundances with an accuracy sufficient for
Galactic studies, temperatures (and hence extinction) and luminosities have to
be accurately determined as well.

In the following subsections we outline the implications of the {\Gaia}
photometric system performance for addressing the science case. However, an
in-depth discussion is not intended. In addition, we comment on the photometric
performance for QSO classification. As mentioned in
Sections~\ref{sec:introduction} and~\ref{sec:chroma}, a proper identification of
QSOs will allow the definition of a non-rotating extragalactic reference frame
which is indispensable for astrometric purposes.
For other objects, we refer to \citet{kaempf} who deal with the automatic
parametrization of unresolved binary stars, to \citet{kolka2005} for the study
of performances for emission-line stars and to \citet{cellino} for a discussion
about {\Gaia} photometry and the parametrization of asteroids.

\subsection{Interstellar extinction and effective temperature}
\label{sec:extinction}

The determination of the effective temperatures, absolute magnitudes and
chemical abundances requires accurate estimates of interstellar extinction 
including the interstellar extinction
law. Ideally the extinction estimates for individual stars are based solely on
each star's observables, as the extinction varies on small scales in any field
except for the closest stars. Assuming a standard extinction law, \Gaia's
photometry allows the determination of such individual extinction measures (see
Figs.~\ref{fig:sigma_post1}, \ref{fig:sigma1} and \ref{g15res}),
though with significantly larger error for the late-type stars due to degeneracy
in the {\av} and {\teff} determinations.  Meanwhile, the determination of
{\teff} for the late-type main-sequence stars remains reliable, even in the
presence of significant extinction, to at least $G=18$. Effective temperatures
with similar uncertainties are achievable for red giants in the
halo where the extinction is low.

We can identify the main contributors to the determination of reddening. For
example, the C1M825--C1M965 vs.\ the C1B655--C1M656 (H$\alpha$ index) diagram is
useful for extinction measurements. The C1M825--C1M965 colour index is primarily
meant as a measure of the strength of the Paschen jump, sensitive to stellar
type, while the H$\alpha$ index is nearly reddening free.  For stars where the
H$\alpha$ index is a measure of the strength of the Balmer line, i.e. stars
earlier than about G3--G5, the unreddened main sequence displays a sharp locus in
this colour-colour diagram. The same H$\alpha$ index value may, however, be
measured for stars on either side of the Balmer maximum (A1--A3 stars), but this
ambiguity is lifted by comparison to the C1M656--C1M965 vs.  H$\alpha$ index
diagram. Thus the inclusion of the H$\alpha$ index allows a classical approach
to determining intrinsic colours from standard curves and simultaneously
correcting the colours for the influence of metallicity and evolution across the
main sequence. But again, intrinsic colours only result when the H$\alpha$ index
is measured precisely, while a potential drawback of the C1M825--C1M965 index is
that these two C1M passbands are among those least affected by extinction.

If the reddening in only one colour index is determined, then the extinction in
other passbands would have to rely on an assumed extinction law. Indeed, in our
performance analysis (Section~\ref{sec:phot_performances}) we have assumed a
standard extinction law. Deviations from this standard law \citep{fitz}, due to
spatial variations in the chemical composition and size distributions of
interstellar dust, will contribute to systematic errors in the individual
extinction determinations. The breadth of \Gaia's photometric system justifies
addressing the question of whether the extinction law itself (averaged over the
line-of-sight to a star) can be determined from the photometry.
Since {\Gaia} is a comprehensive survey of all
point sources, the number of sources is large enough that most stellar types are
sufficiently represented even in small fields, close to the Galactic plane,
where extinction correction is most essential. In this case, local extinction
laws may possibly be deduced from colour difference vs.\ colour diagrams. For
example, deviant extinction laws may be identified from $Q$-like indices vs.\
colour diagrams.

{\Gaia} RVS observations will provide other measures of extinction. First, there
is the diffuse interstellar band at 862~nm located within the wavelength range
of the RVS instrument \citep{RVSI}. In addition, the equivalent width of the IR
calcium triplet gives a rough indication of the \teff; in combination with the
virtually reddening free C1M395--C1M410 index the calcium triplet can be used
as a diagnostic of intrinsic colours and thus to estimate the
extinction. Together with {\Gaia} parallaxes, all of these measures of
extinction (spectroscopic and photometric) will allow the construction of a 3D
Galactic extinction map.

\begin{table*}
\caption{Distances for which the relative parallax error is $\sim$10\%: $d_{\rm
o}$ is the value of this distance for zero interstellar extinction and $d_{\rm
abs}$ is the value for an average Galactic plane interstellar extinction of 0.7
mag kpc$^{-1}$.  $V(d_{\rm o})$ and $V(d_{\rm abs})$ are the corresponding
apparent $V$ magnitudes. Parallax accuracies are from Table 8.4 in
\citet{gsr}}
\label{distances}
\begin{center}
\begin{tabular}{lccccccccccc}
\hline
SP &$M_V$&  $d_{\rm o}$(pc)&$V(d_{\rm o})$& $d_{\rm abs}$(pc)&$V(d_{\rm abs})$&
SP &$M_V$&  $d_{\rm o}$(pc)&$V(d_{\rm o})$& $d_{\rm abs}$(pc)&$V(d_{\rm abs})$\\
\hline
B1 V &$-$3.2& 20 000 & 13.2  & 7000&15.7  &G8 III&0.8  &  9000 &15.6 & 4400&17.1 \\
A0 V &0.65&   8500 & 15.2  & 4500&16.8  &K3 III&0.3  & 10 000&15.3 & 4800&17.1 \\
A3 V &1.5 &   7000 & 15.7  & 3800&17.1  &M0 III&$-$0.4 & 13 000&15.2 & 5500&17.2 \\
A5 V &1.95 &  6500 & 16.0  & 3500&17.3  &M7 III&$-$0.3 & 17 000&15.9 & 6300&18.1 \\
F2 V &3.6 &   4500 & 16.7  & 2700&17.8  \\
F8 V &4.0 &   4000 & 17.0  & 2500&18.1  &B0 Ib &$-$6.1 & 33 000&11.5 & 9500&15.4 \\
G2 V &4.7 &   3500 & 17.2  & 2200&18.2  \\
K3 V &6.65&   2400 & 18.4  & 1700&19.1  &WD    & 8.0 & 1500  &18.9 & 1200&19.2\\
M0 V &8.8 &   1500 & 19.7  & 1200&20.0  \\
M8 V &13.5&   500  & 21.8  & 450 &22.1 \\
\hline
\end{tabular}
\end{center}
\end{table*}

{\Gaia} measurements will also allow a statistical approach to the construction
of a 3D extinction map. Given the parallax, main sequence loci may be
empirically constructed from nearby, unreddened samples provided by \Gaia.
Shifting these loci to a given distance with a given mean extinction identifies
the main sequence in the observed colour-magnitude diagram (including the
photometric errors) for a sample of main sequence stars selected from a small
volume element in the Galaxy.  Thus, main sequence fitting to such stellar
samples, selected by means of the {\Gaia} parallaxes, could serve as the basis
for a 3D extinction map \citep{knude2002}. The chemical composition of the stars
may be used to refine the definition of an appropriate main sequence, and
{\logg} to refine the selection of an appropriate stellar sample.  This approach
has the further advantage that the C1B passbands may be used in addition to the
C1M ones, and that stars along the entire main sequence will contribute to the
estimation of extinction, including those located on the lower main sequence
where extinction determination is otherwise difficult.

The combination of {\Gaia} and external IR photometric data provides independent
determinations of extinction. During the luminosity calibration of 2MASS
photometry for A9--G5 main sequence stars, \citet{knudeclaus} noticed
significant reddening vectors in the $M_J$ vs.  $(J-H)_\mathrm{obs}$ diagram
where $M_J$ was estimated from \textit{Hipparcos} parallaxes $\pi >
8.0$~milliarcsec and $\sigma_{\pi}/\pi <0.11$, assuming that reddening was
negligible. These vectors are most pronounced for the A type stars but are also
present for cooler ones.  From these vectors the extinction law may be estimated
even for various spectral classes if the stellar density is large enough.
Similar diagrams may be constructed from {\Gaia} astrometry and BBP and MBP for
the optical region and for the near-infrared from the 2MASS, UKIDSS and VISTA surveys
so that regions with abnormal extinction can be identified.

\subsection{Absolute luminosity and gravity}
\label{sec:luminosity}
                                                                                                   
Photometry will be crucial for absolute luminosity (or gravity) determination
when relative parallax errors are larger than 10--20\%. The distances at which
the relative parallax error is $\sim$10\% are listed in Table~\ref{distances} as
a function of spectral type and luminosity class in the absence of
interstellar extinction, i.e.\ similar to the Galactic pole direction, and in
the case of an average Galactic plane interstellar extinction of 0.7 mag
kpc$^{-1}$.
The parallax accuracies are from Table 8.4 in the {\Gaia} Study Report
\citep{gsr}. Based on the Besan\c{c}on Galaxy model \citep{robin_gal}, some
100--200 million stars are predicted to have parallaxes measured at the 10\%
accuracy level.

Absolute magnitude calibrations can be established using the stars with
accurate parallaxes and interstellar extinction, and applied to more
distant stars in the classical way, assuming that these are intrinsically
similar to the accurate-parallax stars, which may not always be the case.
Thus luminosities with precision $\sigma_{M_V}\sim0.2$--$0.4$~mag (or
$\sigma_{\logg}\sim0.1$--$0.2$~dex) from photometry are desirable to
match the uncertainties of the well measured parallaxes.

The main contributors to the photometric luminosity determination are the
measurements of Balmer and Paschen jumps, the H$\alpha$ index, the
C1M395--C1M410 vs. W(CaT*) diagram and C1M515 combined with contiguous
pseudo-continuum passbands, as already noted in Section~\ref{sec:BBPMBP}.

Figures~\ref{fig:sigma_post1}, \ref{fig:sigma1}, and \ref{g15res}
show that $\sigma_{\logg}\sim0.1$--$0.2$~dex is achievable for giants and early
type stars at large distances and therefore that the study of Galactic structure
on large scales (warp, spiral arms, outer halo, etc) is feasible.  For nearby
stars the photometric determination of {\logg} is also possible but of less
interest because of the good parallax precision. For distant late type dwarf
stars, i.e.\ with parallax errors larger than 10\%, the precision of the
photometrically determined {\logg} will be not as good as for giants and early
type stars, as expected.

\subsection{Chemical abundances}
                                                                                                   
High resolution, high SNR spectroscopic observations allow the most
reliable chemical abundances determinations \citep[e.g.,][] {cayrel}.
RVS spectra will be used to determine the atmospheric parameters, when possible.
\citet{RVSI} estimate $\sigma_{\feh}$ to be smaller than $0.2$--$0.3$~dex for
stars with $V\sim14$ and to improve by combining the spectra with photometric
and astrometric data. Thus, atmospheric parameters for 10--25 million stars to
$V\sim14$--$15$ will be determined including individual abundances for 2--5
million stars to $V\sim12$--$13$ (mainly Fe, Ca, Mg and Si for F-G-K stars ; N
in hotter stars, such A-type stars; and information on C, N or TiO abundances
for cool K and M-type stars).

Although 10--25 million stars with RVS chemical abundance determinations
may seem little compared to the 1 billion objects observed by \Gaia, the
spectroscopic chemical compositions are crucial as they will serve to
calibrate the photometric data. This is important because for the vast
majority of stars, chemical abundances will be derived from photometric
data exclusively.

The chemical abundances, and especially the relation between {\feh} and
{\afeh}, are indicative of the initial star formation rate (SFR) and
provide the rate of chemical enrichment of the Galaxy. Calculations by
\citet{maeder} show that the $\alpha$-element abundance ratio starts to
decrease at high {\feh} as the initial SFR increases. In agreement with
this, \citet{nissen} argues that the thick disk stars underwent chemical
evolution with a high initial SFR such that a relatively high metallicity
($\feh\sim-0.4$) was reached before the contribution by iron due to the
SNe~Ia decreased the {\afeh} value.  He also argues that the initial SFR
in the thin disk has been slower and that the relative $\alpha$-element to
iron abundances started to decrease at lower metallicity ($\feh\sim-0.6$).
Most of the halo stars were formed at an even lower initial SFR, so the
decrease of {\afeh} occurs at low metallicity ($\feh\sim-1.2$). Some of
the halo stars have {\afeh} abundances typical of the thick disk which
points to a dual model for the halo formation: an inner part which had a
high SFR and an outer part which experienced a slower evolution or was
accreted from dwarf galaxies.

Figures~\ref{fig:sigma_post1}, \ref{fig:sigma1}, and \ref{g15res}
show that {\Gaia} photometry is able to match the expected spectroscopic
precision up to about 1--2~kpc from the Sun depending on the Galactic direction
(i.e.\ the reddening) for F--M dwarfs and subdwarfs. For giants in the red clump
and the red giant branch, both being brighter, the same precision is attainable
up to about 4, 7 and 12~kpc in the centre, anti-centre and orthogonal Galactic
directions, respectively. This implies that the following issues (and other
goals) can be addressed: the determination of the Galactic chemical abundance
gradient, the classification of the stars into different stellar populations,
the characterization of halo streams,
and the determination of the distance scale through the metallicity
determination for RR-Lyrae.

As expected, low metallicities are determined with less precision than
solar or higher metallicities.  Extensive spectroscopic follow-up efforts
from the ground will be necessary to determine the metallicities of stars
with $\feh<-3.0$ with a sufficient accuracy to select the most interesting
targets for high-resolution spectroscopy (Christlieb, private
communication).

Photometric chemical composition determinations are not free of difficulties.
For instance, cool unresolved binaries tend to mimic a single star with a lower
metal content. As an example, for G and K dwarfs with companions for which
$\Delta m$ is $1.5$--$2$ mag the systematic metallicity error is $\Delta\feh\sim
-0.4$. This bias decreases to $\Delta\feh\sim -0.2$ when $\Delta m=3$.

As discussed in Sec.~\ref{sec:phot_performances}, the first trials to determine
{\afeh} using the NextGen2 library ($\feh\in[-2,+0]$ and $\afeh\in[-0.2,+0.8]$)
show that precision of 0.1--0.2~dex for $G\sim15$--$16.5$ can be achieved.  The
determination may be improved with the inclusion of parallax information,
because it constrains the value of {\logg}, which may break the degeneracy in
the variation of the Mg Ib triplet due to luminosity and Mg abundance changes.
For stars fainter than $G\sim16$--$17$, deriving $\alpha$-elements abundances
from photometry is unlikely.

\subsection{Age}
\label{sec:age}

The location of a star in the HR-diagram does not allow for a unique age
determination as several combinations of chemical composition (\feh, \afeh, $Y$,
etc) and age are possible \citep[see e.g.,][]{binney}. Individually accurate
ages require extremely accurate determinations of luminosity, temperature,
reddening and chemical composition, which {\Gaia} will provide.

For ages around 10--14~Gyr, and assuming a given chemical composition, a change
of $\log\teff$ by $\sim0.01$~dex for the stars in the turn-off translates to an
age variation of 2~Gyr.
Thus, temperatures must be known with precisions better than a few hundredths in
$\log\teff$, i.e.\ $1.5$--2\% in \teff, for the turn-off stars to determine
individually accurate ages. A variation of 0.3~dex in {\feh} is equivalent to a
change of $\log\teff$ by $\sim0.01$~dex for a given age. The variation of the
$\alpha$-elements abundance has a smaller impact on the age determination than
the variation of \feh. A variation of $+0.3$ dex in {\afeh} leads to a variation
of $-0.006$ dex in the temperature turn-off, yielding an uncertainty of about
1~Gyr in the age.

In summary, an uncertainty of about $4-5$~Gyr is estimated for the individual
ages of the stars at the turn-off for the halo, thick and old-thin disk stars up
to about 2, 3 and 5~kpc in the Galactic centre, anti-centre and orthogonal
directions, respectively, as deduced from the PS performances in
Section~\ref{sec:phot_performances}.

The age determination for F-G sub-giant stars is quite insensitive to
uncertainties in {\teff}
%, as pointed out by \citet{edvardsson}, 
because the isochrones are almost horizontal. An uncertainty in $M_V$ of about
$0.15$ translates to an uncertainty in the age of about 2~Gyr. Assuming
additional uncertainties in {\feh} and {\afeh} determinations of about $0.3$
dex, the final precision of the individual ages is about $3-4$~Gyr.

\citet{Kucinskas} showed that early-AGB stars can provide ages as accurate as
the turn-off stars for $\feh>-1.5$ when $\sigma_{\log\teff}\sim 0.01$,
$\sigma_{\logg}\sim 0.2$, $\sigma_{\feh}\sim 0.2$ and $\sigma_{E_{B-V}}\sim
0.03$, thus allowing the age determination to larger distances than for turn-off
and sub-giant stars. The {\Gaia} capabilities in the case of metal-poor stars
have not yet been investigated.

Subsets of each Galactic population (such as globular clusters, open clusters,
OB associations, a given halo stream, an identified merger, a moving group,
etc.) can be treated statistically and mean ages and chemical abundances of the
group can be obtained with much better precisions than those of the individual
members.

The prospects for the determination of the age-metallicity relation and the star
formation history with {\Gaia} are discussed by \citet{misha}.  Assuming the
errors on the astrophysical parameters from Section~\ref{sec:phot_performances},
the author recovers the simulated age metallicity relation in the case of
moderate extinction and for a slowly varying star formation rate. The details of
the star formation history, such as a series of prominent short bursts, are not
recovered.

\subsection{QSOs}
\label{sec:qso}

QSOs play an important role in the {\Gaia} mission as they will be used to
construct the astrometric reference frame and they are of course interesting in
their own right. However, since the QSO population only represents 0.05\% of the
stellar population, building a {\it secure} QSO sample with no stellar
contamination requires a very efficient rejection algorithm. Although proper
motion, parallax and variability information will help in rejecting stars, these
will be available with the required precision only at the end of the mission.
Therefore it is important to check the capability to classify QSOs using only
photometric data.

The QSO classification efficiencies of Supervised NNs and of MDMs (see
Sec.~\ref{sec:nn}) have been compared using synthetic data generated from the
BaSeL 2.2 library for stars \citep{lejeune}, from pure-hydrogen atmosphere
models for white dwarfs (Koester, private communication) and from a library of
QSO synthetic spectra \citep{cla1,cla2}. Properly trained NNs are found to be
capable of rejecting virtually {\it all} stars, including white dwarfs.  This is
at the expense of the completeness level of the QSO sample, being only $\sim
20\%$ at $G=20$. However, this provides a sufficient number of objects in the
context of the non-rotating extragalactic reference system determination. MDMs
provide a higher completeness level ($\sim 60\%$ at $G=20$), but with a
correspondingly higher stellar contamination rate of the QSO sample. MDMs may be
preferred at high Galactic latitudes. Not surprisingly, reddened QSOs and weak
emission-line objects are preferentially lost while high red-shift objects are
most easily recognized.

Assuming an object is a QSO, it is also possible to infer its red-shift from its
photometric signature in the C1B+C1M passbands. Unfortunately, there is a colour
degeneracy in the QSO spectra which limits the expected precision of the
technique to about $|\Delta z|_{\rm Median} \simeq 0.2$ in the range $0.5 <
z_{\rm spec}< 2$ (computations were done with a previous version of the {\Gaia}
PS but are not strongly sensitive to the adopted filter sets).

More details on the identification and characterization of QSOs with {\Gaia} can
be found in \citet{cla1}.

%%%%%%%%%%%%%%   Conclusions   %%%%%%%%%%%%%%%%%%%%%%%%%%%%%%%%%

\section[]{Conclusions}
\label{sec:conclusions}

To fully achieve the scientific goals of the {\Gaia} mission it is essential to
complement the astrometric and radial velocity measurements with highly accurate
multi-colour photometry. This is necessary both for obtaining accurate
astrometry and for a proper scientific interpretation of the stereoscopic census
of the Galaxy {\Gaia} will provide.  The latter goal requires that the
photometric system for {\Gaia} is capable of astrophysically parametrizing
1~billion objects across the entire Hertzsprung-Russell diagram in the presence
of varying degrees of reddening and for the full range of chemical compositions
and ages of stars populating the Galaxy, and, of identifying peculiar objects
and QSOs.

There are no existing photometric systems capable of handling this very
demanding task. We therefore set out to design a new system for the {\Gaia}
mission. The broad passbands implemented in the Astro instrument should fulfil
the chromaticity requirements from the astrometric data processing while at the
same time allowing for the astrophysical characterization of stars in very
crowded regions. The main classification and astrophysical parametrization task
is carried out with the medium passbands implemented in the Spectro instrument.

The novel development we introduce in this work is the use of an objective
`figure of merit' (FoM) to compare different proposals for the {\Gaia}
photometric system.  This FoM is based on the predicted errors that can be
obtained for the parametrization of stars in terms of \teff, \logg, \av, \feh,
and \afeh. These errors are calculated by using synthetic spectra to evaluate
the sensitivity of each photometric passband to the astrophysical parameters. At
the same time the degree of local degeneracy in the astrophysical parameters is
taken into account in the FoM. The overall FoM for a photometric system is
calculated taking into account the priorities of the different scientific
targets that were deemed to be most important for addressing the core science
case for \Gaia.  Using the FoM allows us to choose objectively which of the many
proposed photometric systems is best in terms of reaching the astrophysical
parameters error goals while at the same time having the least number of local
degeneracies.

The calculation of the predicted posterior errors and the FoM as outlined in
Appendix~\ref{ap:fom} can be applied to any photometric system and can also be used to predict
parametrization errors that can be achieved with spectra of stars. In addition
our method can be extended to the optimization of systems that can be used for the
study of high red-shift galaxies or quasars (for example to derive photometric
red-shifts), as long as the dependence of the relevant SEDs on parameters such as
red-shift or star formation rate are known.

The main contribution of this work is the photometric system itself. 
This system has been designed based on our astrophysical knowledge and 
experience with ground-based systems and resulted in five (C1B) and fourteen 
(C1M) passbands implemented in the Astro and Spectro instruments, respectively. 
The number of passbands reflects the variety of targets
(all types of stars, QSOs, galaxies, solar-system objects), the required
stellar astrophysical parameters (\teff, \logg, \av, \feh, \afeh) and the need of
breaking degeneracies (changes in two or more parameters may translate to the
same changes in some spectral features but different changes in other ones). In
summary, three of the broad C1B passbands are located to the left of H$\beta$
line, on the H$\alpha$ line and to the right of Paschen jump. The other two
passbands fill the gaps so that full coverage of the whole spectral range of
{\Gaia} observations is provided. A broad passband in C1M, implemented in the
Spectro instrument, provides the measurement of the UV flux at wavelengths
blueward of the Balmer jump. Seven medium passbands are placed on the crowding
of Balmer series in early-type stars, on the CaII H line, on Mg I triplet and
MgH band, on the H$\alpha$ line, on one of TiO absorption bands in cool stars,
and on the strong CN band in R- and N-type stars. An additional passband serves
to measure the flux in the wavelength range covered by RVS. Finally, five
passbands are devoted to the measurement of the pseudo-continuum.

The unfiltered light measured in the Astro and Spectro instruments provides
photometric data in two very broad passbands (from 400 to 1000~nm and from 350
to 1025~nm). The $G$ magnitude from Astro yields the highest SNR among all
{\Gaia} magnitudes, and hence is the most suitable for variability analysis.

End-of-mission magnitude precisions have been estimated following an `aperture
photometry' approach.  A precision of $0.01$~mag is obtained at $V\sim 18$ and
$~16$ for the C1B and C1M passbands, respectively. In the case of the $G$
passband and for stars brighter than $V\sim$16 and $\sim$14 measured with C1B
and C1M, the photometric precision is limited by how accurately the data are
calibrated. Although there is no detailed calibration model, a rough estimation of
the number of involved calibration parameters and the number of available stars
for calibration purposes shows that sub-millimagnitude photometric precisions
are potentially achievable. A PSF-fitting approach shows that it is possible to
deal with stellar densities up to about 2--4$\times$10$^5$ stars/sq~deg to
magnitude 20 with Spectro, if accurate positional information from astrometry is
used.

The performance of the {\Gaia} photometric system with respect to the 
determination of astrophysical parameters for
single stars has been evaluated using the `posterior' errors from the FoM
formalism and using algorithms designed for the astrophysical parametrization of
stars (based on MDM and NN). The differences and limitations have been
discussed.  Both approaches demonstrate that precision goals in
Section~\ref{sec:test_pop} are generally achieved to $G\sim$17--18~mag and that
the precision for each parameter depends on temperature, luminosity, chemical
composition and interstellar extinction, and apparent magnitude, as expected.
Reasonable uncertainties of 0.5--1~dex are obtained even for the chemical
composition of metal-poor stars (\feh$=-4.0$). The determination of {\afeh} to a
precision of about 0.1--0.2~dex is possible down to $G$$\sim$16--17 for low to
intermediate temperature stars. Assuming a standard extinction law, the C1B and
C1M passbands allow the determination of individual extinctions and hence the
determination of {\teff} is reliable, even in the presence of significant
extinction, down to at least $G=18$. By combining photometry and astrometry, a
statistical approach to the construction of a 3D Galactic extinction map is also
possible. {\Gaia} will yield parallaxes with relative uncertainties lower than
10\% for some 100--200~million stars allowing absolute magnitude determinations
across the entire HR-diagram if the extinction is known. The C1B+C1M system is
designed to provide the same luminosity errors (0.1--0.2~dex in \logg) for
giants and early type stars at large distances ensuring the study of the
Galactic structure on large scales. Individual ages of stars at the turn-off of
the halo and old-thin and thick disk at distances up to about 2, 3 and 5 kpc in
the Galactic centre, anti-centre and orthogonal directions, respectively, can be
determined with a precision of about 4--5~Gyr. The same precisions can be
obtained for early-AGB stars, thus allowing the probing of larger distances than
with turn-off and subgiant stars. For subgiants, our estimations yield slightly
lower uncertainties of 3--4~Gyr. Note that we quote uncertainties for individual
ages. For a group of stars the mean age will be determined to much better
precision.

An additional merit of the {\Gaia} photometric system is its ability to
discriminate QSOs from stars and white dwarfs. This will results in a sufficient
number of objects for the definition of the non-rotating extragalactic reference
system for proper motions.

Finally, the performance of the C1B passbands when dealing with the chromaticity
residuals has been evaluated. Our estimates show that the chromatic contribution
to the parallax errors is 0.14--1.4~microarcsec. As these numbers are based on a
worst-case assumption for the WFE and a somewhat simplistic calibration model, a
residual contribution of 1~microarcsec for stars and of few microarcsec for the
QSOs is likely achievable.

In summary, the photometric system C1B+C1M developed by the Photometry
Working Group satisfies the mission requirements and was therefore adopted
as the baseline by the {\Gaia} Science Team. It has since been adopted by
the ESA Project Team as the basis for the formal mission requirements to
the industrial teams participating in ESA's Invitation to Tender for
\Gaia.

The experience from the development of the photometric system, including
selecting the scientific targets and the `figure of merit' approach, puts us in
a good position to rapidly optimize new payload proposals that will come from
the selected industrial contractor in the course of 2006.

%%%%%%%%%%%%%%   Acknowledgements   %%%%%%%%%%%%%%%%%%%%%%%%%%%%%%%%%

\section*{Acknowledgments} 

We would like to acknowledge the contributions by B. Edvardsson, R. Lazauskaite,
T.  Lejeune and J. Sudzius and the collaborative effort of members of the
Photometry Working Group within the {\Gaia} mission preparation framework.  We
also thank P. Hauschildt, I. Brott and F. Castelli for making available to us
SED libraries with $\alpha$-elements enhanced abundances.
The authors acknowledge the funding support from the following countries:
Belgium, Denmark (through the Danish Ministry of Science), Estonia (through the
ESF grants No. 5003 and 6106), France, Germany (through the DLR and the Emmy
Noether Programme of the DFG), Greece (through GSRT), Italy, Lithuania (through
the Ministry of Education and Science), The Netherlands (through the Netherlands
Research School for Astronomy NOVA and The Netherlands Organisation for
Scientific Research NWO), Russia, Spain (through MCYT under contract
PNE2003-04352), Sweden, Switzerland and United Kingdom. For Denmark, Estonia,
Lithuania and Sweden support from the Nordic Research Board [Ref.\ NB00-N030 and
2001-2783-4] is gratefully acknowledged. A. Ku\v{c}inskas acknowledges the
Wenner-Gren Fellowship. Finally, we would like to thank the referee,
M.S.~Bessell, for a rapid response and constructive comments that helped improve
the paper.

%%%%%%%%%%%%%%   References   %%%%%%%%%%%%%%%%%%%%%%%%%%%%%%%%%%%%%%%%%

%%%%%%%%%%%%%%   Appendix   %%%%%%%%%%%%%%%%%%%%%%%%%%%%%%%%%%%%%%%%%

\appendix
\section{Figure of merit}
\label{ap:fom}

The definition and the procedure for calculating a figure of merit for a
photometric system was proposed by \citet{lin_fom} and we provide the details
here.

The achievable errors {\sikpost} for a particular ST $i$ can be calculated
using the so-called sensitivity matrix $\mat{S}_i$. The elements of this
matrix are the partial derivatives $\partial\phi_{ij}/\partial p_k$, where
$\phi_{ij}$ is the (noise-free) normalized flux in filter $j$ (in photon
counts) for ST $i$, and $p_k$ stands for the astrophysical parameter $k$.
These derivatives describe how the flux in each filter changes in response
to a change in astrophysical parameter (AP) $k$. Consider the AP determination 
as a linearized least
squares estimation of $\Delta\vect{p}$, the improvement of the AP vector.  
The observation equation for ST $i$ resulting from the flux measured in
filter $j$ reads:
\begin{equation}
\label{eq:obseq}
\frac{\partial\varphi_{ij}}{\partial p_1} \Delta p_1 + \dots +
\frac{\partial\varphi_{ij}}{\partial p_K}  \Delta p_K = 
\Delta\phi_{ij}\pm\epsilon_{ij}\,,
\end{equation}

where $\Delta\phi_{ij} = \phi_{ij,\mathrm{obs}} - \phi_{ij}(\vect{p})$ is
the difference between the observed and predicted flux and $\epsilon_{ij}$
indicates the flux uncertainty. Observation equations of unit weight are
formed through division by $\epsilon_{ij}$, whereupon normal equations are
formed in the usual manner.  The linearization is assumed to be made
around the true parameter vector \vect{p}, so that the resulting update
$\Delta\vect{p}$ has zero expectation.  Then, given the
variance-covariance matrix
$\mat{C}_{\vecphi}=\mathrm{diag}(\epsilon^2_{ij})$ of the observed fluxes,
the variance-covariance matrix of the estimated AP-vector
$\vect{p}_\mathrm{post}$ is given by the inverse of the normal equations
matrix:
\begin{equation}
\label{eq:cpostnob}
\mat{C}_{\vect{p},\mathrm{post}}=
(\mat{S}^\mathrm{T}\mat{C}_{\vecphi}^{-1}\mat{S})^{-1}\,,
\end{equation}
where the matrices $\mat{C}_{\vect{p},\mathrm{post}}$,
$\mat{C}_{\vecphi}$, and \mat{S} are defined for each ST $i$ separately.
The diagonal elements $[\mat{C}_{\vect{p},\mathrm{post}}]_{kk}=\sikpost$
of this matrix are the sought after achievable errors for a PS. In reality
degeneracy among the APs will often make the matrix
$(\mat{S}^\mathrm{T}\mat{C}_{\vecphi}^{-1}\mat{S})^{-1}$ singular or
near-singular, resulting in infinite or very large {\sikpost} as computed
from equation (\ref{eq:cpostnob}). This can be avoided by adding a suitable
positive definite matrix \mat{B} which makes the whole right-hand side
positive definite; thus:
\begin{equation}
\label{eq:cpost}
\mat{C}_{\vect{p},\mathrm{post}}=
(\mat{B}+\mat{S}^\mathrm{T}\mat{C}_{\vecphi}^{-1}\mat{S})^{-1}\,.
\end{equation}
\mat{B}\ is the \textit{a priori} information matrix of the APs. In the
absence of any other information on the APs we have
$\mat{B}=\mathrm{diag}(\sikprior^{-2})$. This matrix plays an important
role and can be used to incorporate prior information on the astrophysical
parameters. This matrix can also be used to incorporate constraining
information from the parallax measurements.  When the photometric data
does not provide any relevant information on a given AP $p_k$ (either
because the flux variances in $\mat{C}_{\phi}$ are too large or because
the elements of the sensitivity matrix $\mat{S}$ are too small), then
$\sikpost\simeq\sikprior$.

In practice the derivatives $\partial\phi_{ij}/\partial p_k$ are
calculated numerically from simulated photometric data. The calculation
thereof requires (synthetic) spectral energy distributions (SEDs) of the
STs and a noise model for the photometric instruments.

The figure of merit is now calculated as follows. For each ST $i$ and
astrophysical parameter $k$ (\av, \feh, \logg, \teff, \afeh, \dots) the
performance of the photometric system is measured by the ratio
$\sikpost/\sikgoal$. The figure of merit $Q_i$ for each ST $i$ is then
defined as:
\begin{equation}
\label{eq:calcqi}
Q_i=\sum_k w_k f(\sikpost/\sikgoal)\,,
\end{equation}
where $w_k$ indicates the relative weight of each astrophysical parameter
(with $\sum_k w_k=1$) and $f(x)$ is a non-linear function of
$x=\sikpost/\sikgoal$ with a break around 1. The function used is:
\begin{equation}
\label{eq:fx}
f(x)=(1+x^{2n})^{-1/n}\,.
\end{equation}
The global figure of merit (summed over all STs) is the weighted and
normalized sum:
\begin{equation}
\label{eq:calcq}
\qnorm=\frac{\sum_i w_i Q_i}{\sum_i w_i}\,,
\end{equation}
where the weights $w_i$ indicate the priority of each ST $i$. The value of
\qnorm\ indicates how close the performance of a photometric system is to
being `ideal' (when $\qnorm=1$). This global figure of merit is then
calculated for each photometric system using the error goals, weights and
priorities defined in \citet{jordiprio}.

The value of $n$ in the function $f(x)$ from equation (\ref{eq:fx}) determines
how much weight good performance ($x\le1$) gets in the FoM as opposed to
bad performance ($x>1$). We are looking for a PS that achieves the error
goals over all of AP space and we want to avoid giving a high rank to a
system that is extremely good in only one corner of AP space but very bad
everywhere else. The latter will be highly ranked for $n = 1$ but not for
large $n$. However, for very large $n$ (for which $f(x)$ approaches a step
function with $f(x)=0$ for $x>1$) we will not be able to tell the
difference between two PSs that are equally bad wherever $x > 1$ but with
one PS being better than the other when $x < 1$. We chose the value of $n
= 3$ as a good compromise between these two extremes.

We end with a few remarks about the sensitivity matrix. The columns of
$\mat{S}_i$ are the gradient vectors that describe the changes of the fluxes
with respect to a change in an AP. Thus $\mat{S}_i$ contains all the information
needed to characterize the behaviour of the photometric system in the data space
(or filter flux space) near ST $i$. In the ideal case the gradient vectors
would be aligned with the coordinate axes of the data space. That is each flux
measurement $\phi_{ij}$ is sensitive to one (and only one) AP. In practice the
gradient vectors are of course not aligned with the coordinate axes which means
that each AP influences some linear combination of fluxes. This also means that
even if all the gradient vectors are orthogonal the errors in the APs will still
be correlated because any error on a measured flux $\phi_{ij}$ will influence
multiple AP determinations. These correlations are correctly taken into account
in the calculation of the variance-covariance matrix of the estimated AP-vector
and are reflected in non-zero off-diagonal elements in this matrix.

A further complication that will occur in practice is that the gradient vectors
will not be orthogonal to each other. This means that there will be degeneracies
between the APs when we try to estimate them. A well-know example is the
degeneracy between {\teff} and {\av} if only the continuum of the spectrum is
measured. The behaviour of the PS depends on the noise and we should in fact
consider the noise-weighted gradient vectors which have components
$1/\epsilon_{ij}\times\partial\phi_{ij}/\partial p_k$. When the orthogonality
is defined with respect to the noise-weighted gradient vectors, non-orthogonal
gradient vectors will lead to larger correlations between the errors in the
estimated APs \textit{and} they will also cause the standard errors on each AP to
increase. Without going into the mathematical details this can be appreciated if
one considers that any degeneracy between two APs will make it more difficult to
attribute flux changes to either of them thus increasing the uncertainty in both
parameters.

This means that a PS that contains larger degeneracies will get a lower
figure of merit because of the increased {\sikpost} for degenerate APs.
Note, however, that the FoM used in this sense is a measure of how good a
particular PS is at \textit{locally} separating stars with different APs
along orthogonal directions. The FoM does not take \textit{global}
degeneracies into account, where very different parts of AP space are
mapped onto each other in filter flux space.

\label{lastpage}

\end{document}